\documentclass[aps,prx,twocolumn]{revtex4-1}

\usepackage{amsmath}
\usepackage{amsthm}
\usepackage{amsfonts}

\usepackage{bbm,graphicx,bm}
\usepackage[colorlinks=true,
  urlcolor=blue,
  linkcolor=blue,
  citecolor=blue]{hyperref}

\usepackage{color}

\makeatletter
\def\toc@secskip{0.4em}
\def\l@f@section{%
 \addpenalty{\@secpenalty}%
 \addvspace{\toc@secskip plus\p@}%
}%
\def\appendix@toc{%
 \addvspace{0.4em plus\p@}%
 \def\toc@secskip{0.05em}%
}%
\makeatother

\makeatletter
\newcommand*{\defeq}{\mathrel{\rlap{%
                     \raisebox{0.3ex}{$\m@th\cdot$}}%
                     \raisebox{-0.3ex}{$\m@th\cdot$}}%
                     =}
\makeatother
\def\ave#1{\langle #1 \rangle}
\def\ii{{\rm i}}
\def\sx{\sigma^{\rm x}}
\def\sy{\sigma^{\rm y}}
\def\sz{\sigma^{\rm z}}

\def\hx{h_{\rm x}}
\def\hz{h_{\rm z}}

\def\1{\mathbbm{1}}
\def\tr#1{{\rm tr}(#1)}
\def\ket#1{{| #1 \rangle}}
\def\bra#1{{\langle #1 |}}
\def\braket#1#2{{\langle #1 | #2 \rangle}}
\def\bracket#1#2#3{{\langle #1 | #2|#3 \rangle}}
\def\l2{{\ell_2}}
\def\nor#1{\|#1\|}
\def\e#1{{{\rm e}^{#1}}}
\newcommand{\abs}[1]{\left\lvert#1\right\rvert}

\def\M{{\cal U}}
\def\Mr{{\cal U}_r}
\def\Aa#1{A^{(\boldsymbol{#1})}}
\def\a#1{a^{(\boldsymbol{#1})}}
\def\ba{\boldsymbol{\alpha}}
\def\bb{\boldsymbol{\beta}}
\def\R{{\mathfrak R}}
\def\L{{\mathfrak L}}
\def\P{{\cal P}}
\def\dt{d_{\rm T}}
\def\dx{d_{\rm s}}
\def\vv{{v_{\rm B}}}

\def\tit#1{{\em #1},}

\newcommand{\dif}[1]{{#1}}
\newcommand{\difa}[1]{{#1}}
\newcommand{\mz}[1]{{#1}}

\begin{document}

\title{Quantum many-body operator cascade as a route to chaos}

\author{Urban Duh}
\email{urban.duh@fmf.uni-lj.si}
\affiliation{Physics Department, Faculty of Mathematics and Physics, University of Ljubljana, 1000 Ljubljana, Slovenia}

\author{Marko \v Znidari\v c}
\affiliation{Physics Department, Faculty of Mathematics and Physics, University of Ljubljana, 1000 Ljubljana, Slovenia}

\date{\today}

\begin{abstract}
  \dif{We describe a chaos scenario valid for chaotic many-body quantum
  systems, the simulation of which is one of the prime candidates for
  demonstrating quantum advantage, and clarify its importance for classical
  truncation-based simulations. By studying the spectral properties of the
  truncated propagator, we, for the first time, identify fractal structures
  responsible for many-body quantum chaos, analogous to the fractality in
  classical chaos emerging from the time evolution of initially smooth
  long-wavelength densities into ever finer short-wavelength densities.} Namely,
  we show that \dif{the leading slowest-decaying Ruelle-Pollicott eigenoperators} have a non-trivial fractal dimension quantifying
  their non-locality, visible also in the divergence of their condition numbers.
  Furthermore, we find that unitarity imposes an asymptotic equality between the
  temporal decay rate of local correlations and this many-body fractal
  dimension. With this insight, the scenario for many-body quantum chaos becomes
  clear: over time, local operators evolve toward increasingly non-local ones
  with a quantifiable fractal structure, thereby naturally leading to
  effective non-unitary relaxation on the subspace of local operators -- a kind
  of many-body Kolmogorov cascade in the space of operators -- \difa{which enables
  accurate truncation-based simulations}. \dif{Our predictions are
  demonstrated numerically in various one- and two-dimensional quantum circuits,
  with and without conservation laws, as well as by exact results in
  dual-unitary circuits.}
\end{abstract}




\maketitle
{
\hypersetup{linkcolor=black}
\tableofcontents
}

\section{Introduction}

{\bf Quantum computation.--}
\dif{The main goal of quantum computers is to \difa{outperform the best
classical algorithms on certain tasks} -- achieving the so-called quantum
supremacy~\cite{Qadvantage}. One of the prime
candidate problems is the simulation of quantum many-body systems~\cite{manin,Feynman},
specifically for questions that are beyond the reach of efficient classical methods.
Namely, it is expected that simulating chaotic quantum many-body systems on
classical computers is generically exponentially difficult both in time and
in their size. This is partly based on the analogy with classical chaotic systems,
where, as described in the next paragraph, a core reason for the high complexity
is well understood: time evolution produces fractal structure on ever finer
spatial scales. Indeed, a number
of experiments on quantum computers target the simulation of chaotic quantum many-body
systems~\cite{googlescrambling21,IBMutility23,IBMDU26,IBMint25,feig25,andersenXY}.
In order to demonstrate quantum supremacy by simulating many-body systems, one
must understand two issues: (i) to begin with, what quantum many-body chaos is
and does it possess features similar to those of classical chaos, and (ii) how the best
classical algorithms perform on such models and in which special situations they
nevertheless perform competitively. That is to say, several supremacy claims have
been quickly challenged by ingenious classical simulations, see, e.g.,
Ref.~\cite{google19} and Ref.~\cite{pan22}, or Ref.~\cite{IBMutility23} and
Refs.~\cite{begusic24,miles24}.

In the present paper we clarify both issues.
First, we address an old open question of what quantum many-body chaos is by
identifying fractal structures in the Hilbert space operator dynamics that are
rather analogous to fractal structures found in phase space of classical chaotic
systems. Second, our underlying technical tool, the truncated propagator,
explains when classical simulation techniques based on propagating Pauli
strings~\cite{tibor22, begusic, pauli_prop} work well. Furthermore, we put the
method in a wider context of Ruelle-Pollicott (RP) resonances and show that, for
a specific implementation that we use, it has a rigorous underlying framework in
terms of rigged Hilbert spaces, which allows us to exploit its capabilities
better. We also note that the truncated propagator is a tool
explicitly geared towards observables directly measured in quantum experiments
like correlation functions and their higher moments~\cite{google}. Because the
native mode of many of today's quantum computers is digital rather than analog,
we focus on the dynamics of quantum many-body circuits in one and two spatial dimensions.

Let us first briefly remind ourselves what classical chaos is and why it is hard
to simulate, then what common definitions of quantum chaos are and why they are
not fully satisfactory in the many-body setting relevant for modern quantum
experiments.}

\begin{figure*}[t!]
  \centerline{\includegraphics[width=0.8\textwidth]{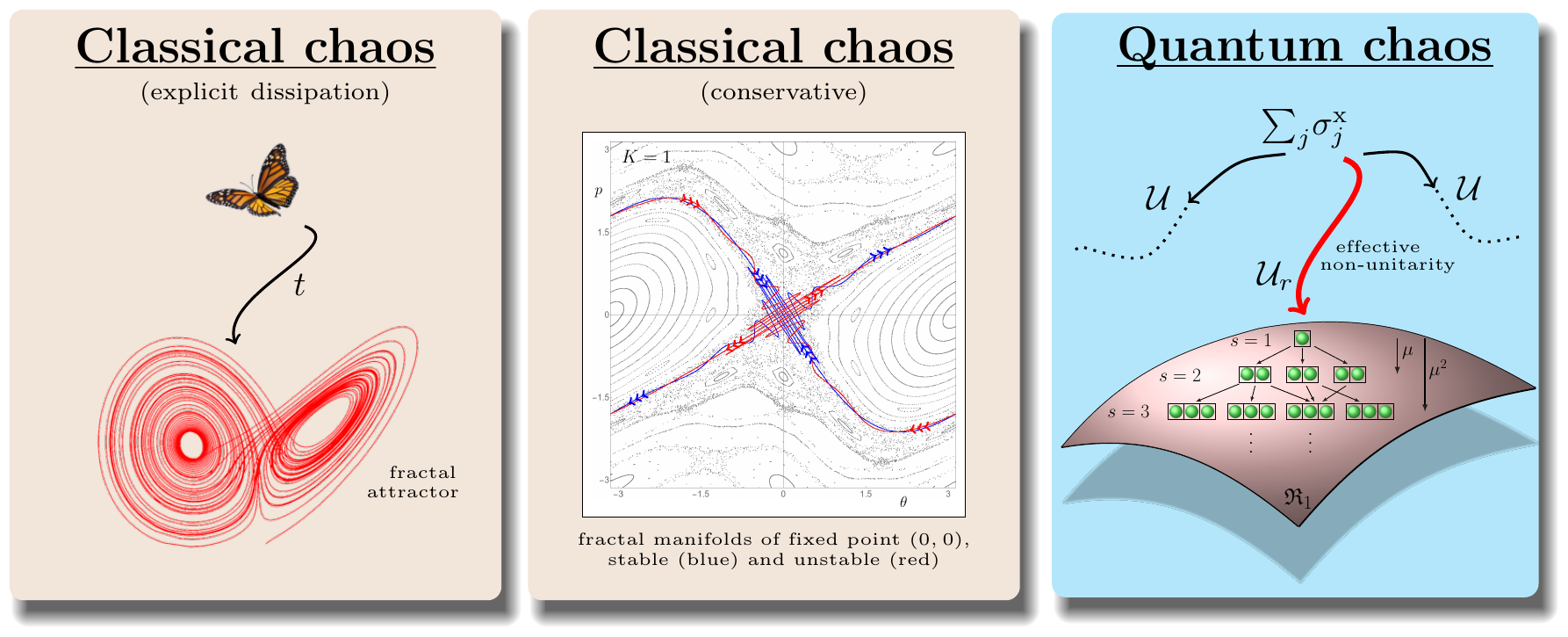}}
 \caption{{\bf Chaos scenarios.} \dif{Classical: Orbits in classical chaotic
 systems with an explicit dissipation converge towards a fractal attractor
 (left, Lorenz system). In conservative systems one has fractal stable and
 unstable manifolds (middle, standard map). Quantum many-body (this paper,
 right): In unitary lattice systems parts of the initial local operator with
 time converge to a manifold $\R_1$, leading to effective non-unitarity
 described by $\Mr$. The manifold has fractal structure where the weight of
 operators on the next level of non-locality $s$ is larger by a scaling factor
 $\mu$.}}
   \label{fig:diagram}
\end{figure*}
{\bf Classical chaos.--}
Understanding of classical chaotic systems has greatly benefited from simple toy
models, for instance, area-preserving maps like the cat map~\cite{Arnold}, in
which one can in detail understand how chaos arises and which then serves to
explain similar core features found also in more complicated chaotic
systems. In conservative chaotic systems, one of the mechanism leading to chaos is
the stretch-and-fold scenario~\cite{Ott,Lichtenberg}; chaotic systems exhibit
stretching (instability) along some directions but must at the same
time also \dif{contract in complementary directions in order to preserve the
\difa{phase-space volume}}. \dif{A small phase-space patch is, therefore, first stretched into a
spaghetti-like object which is then folded back.} Folding is thus a necessary
consequence of unitarity (Liouville symplectic structure) in a finite-volume
phase space. \dif{Phase space of chaotic systems is populated by a dense set of
{\em fixed points} (periodic orbits).} \dif{Unstable fixed points have their
associated stable and unstable manifolds~\cite{foot1} with fractal structure}
\dif{(Fig.~\ref{fig:diagram}, middle): unstable and stable manifolds exhibit an
infinite series of homoclinic intersections around which the local dynamics can
be described by the above-mentioned stretch-and-fold mechanism as, e.g., described
by the Smale's horseshoe map~\cite{Ott}. Those fractal structures form the}
skeleton of chaos~\cite{chaosbook}. In particular, if one starts with a
smooth initial density having a nonzero overlap with any unstable fixed point, the
time evolved density will increasingly concentrate along the unstable
manifold, becoming more and more fractal-like with increasingly smaller
features. Relaxation, being one of the most important properties of chaotic
systems, therefore appears because the ``information'' flows from large to
infinitely small spatial wavelengths enabled by fractal structures of stable and unstable manifolds. \dif{Such classical chaos is
well understood in systems with few degrees of freedom (in middle
Fig.~\ref{fig:diagram}, we show a one degree of freedom system having
2-dimensional phase space), while in systems with many degrees of freedom, e.g.,
many-particle classical systems, details involved are
less clear.

Infinitely small spatial features of classical chaos are enabled by an
infinite-dimensional functional space of densities. As we shall see, this is an
important remark -- true chaos, classical or quantum, which is defined in the limit $t \to \infty$, can
only appear in an infinite-dimensional functional space of some kind. While in classical systems the functional space is \difa{usually} infinite dimensional
due to an ``infinite spatial resolution'', another possibility to get an infinite
space is sending the number of degrees of freedom to infinity. This is precisely
the case that we will study in the quantum domain.

Let us also mention classical dissipative systems, for which the chaos scenario
is a bit different, however, one can again identify fractal features. In
particular, with time, orbits converge to a lower-dimensional fractal subset
called the strange attractor. \difa{As explained in the previous paragraphs,
unstable manifolds play a similar role in conservative systems, since
phase space densities increasingly concentrate on them with time. In fact,
the evolution of phase space densities to ever more fractal unstable manifolds
can be interpreted as effective dissipation in purely unitary dynamics.
One of the main results of the present paper is the identification of analogs
of attractors/unstable manifolds in many-body quantum systems.}}

{\bf Quantum chaos.--}
In quantum systems that have a well-defined classical limit, one can play a
similar game as in classical systems by relying on the semiclassical limit
(e.g., high energies,
or, formally, $\hbar \to 0$) in which the Hilbert space becomes infinitely
large. However, in this work we are interested in quantum systems without any
classical limit, like lattice systems of \dif{qubits which are of interest in
quantum information~\cite{someQC}. Such systems have a finite-dimensional local
Hilbert space (i.e., finite $\hbar$) as well as a finite spatial lattice spacing,
so no classical-like chaos that requires infinite spatial resolution is
possible. The only possibility to get an infinite total Hilbert space, and
therefore be able to even speak about chaos, is to have infinitely many qubits
(degrees of freedom), that is, to take the thermodynamic limit (TDL). This is
precisely the limit of interest in quantum computation -- large systems -- and
this is why modern quantum computer experiments and quantum chaos are so
intertwined. The question, though, still is:} is there any fractality to be found
which could then in turn be used to characterize quantum chaos\dif{, giving
support to the thesis that quantum chaotic systems are similar to classical
chaotic ones, e.g., are hard to classically simulate}?

\dif{In analogy with classical systems, a natural approach might} be to study
fixed points of an appropriate quantum propagator $U$ or a Hamiltonian $H$,
i.e., properties of its eigenvectors and eigenvalues. This is indeed a very
fruitful historical approach in single-particle quantum
systems~\cite{Haake,gutzwiller} in which one studies, e.g., the level spacing
statistics or the eigenvector properties, for instance the eigenstate
thermalization hypothesis~\cite{ETHrev}. In quantum circuits in the
TDL though, there are problems with such an approach.
Namely, the Hilbert space in the TDL is infinitely-dimensional, and the spectrum often becomes
continuous, in line, in fact, with the necessary spectral properties
of a truly chaotic system~\cite{gaspard} (or, more precisely, of a mixing
system). Even more pertinent, the eigenvectors do not exist for the continuous
part of the spectrum~\cite{functional}, although they often can be defined by
extending the considered Hilbert space~\cite{gelfand}. \dif{Those
``theoretical'' difficulties are reflected also experimentally -- many-body
eigenstates and eigenvalues are not really measurable quantities (e.g.,
exponentially close eigenstates would require exponentially long time to be
resolved)}. For that reason, the whole process of studying spectral properties
of finite matrices and then increasing their size has to be approached with
utmost care.

{\bf Many-body quantum chaos.--}
Hints of how to remedy the above deficiencies come from both mathematics and
physics: In functional analysis, one must take into account the space on which
operators are defined, while from the physical point of view one expects that
only local enough observables relax, whereas the non-local ones, for instance an
eigenprojector, do not. \dif{Rather than studying unobservable many-body
eigenstates, let us focus on measurable observables.}

The relaxation rate of (local) observables can be extracted using the
so-called Ruelle-Pollicott formalism~\cite{Pollicott85,Ruelle86,gaspard},
in which the singularity $\lambda_1$ (the RP resonance) of the resolvent inside
the unit circle, or, equivalently, the singularity of the
correlation function in the complex frequency plane, determines the asymptotic
decay of dynamical correlations,
\begin{equation}
  C(t)=\langle A(0)A(t)\rangle \sim c_1\lambda_1^t,
\end{equation}
where $A$ is a physical observable (with local density) and the expectation value is taken with
respect to some invariant measure $\rho$. We will always consider
infinite temperature $\rho \sim \1$.
There are different ways to extract $\lambda_1$; in this work we will follow a physical
intuition and only briefly comment on the mathematical approach of rigged Hilbert
spaces leading to the same picture \dif{(for details on that see Ref.~\cite{CMV})}. The main idea is that one should study the spectral properties of the propagator truncated to the space of local
operators~\cite{Prosen}, that is, focus on operators that are of physical
interest. As we are going to show, spectral properties of the so-called truncated
propagator $\Mr$, in contrast to those of unitary $U$, do carry well-defined physical
information even in the limit of an infinite system size.

\subsection{Summary of results}

Taking the unitary propagator of operators in an infinite homogeneous
lattice system (i.e., working in the TDL) and truncating it to a subspace
of operators with their density supported on at most $r$ consecutive sites, we
obtain the truncated propagator $\Mr$. In the limit $r \to \infty$, $\Mr$
accurately describes the dynamics of local operators, for instance, the
relaxation of correlation functions. In chaotic systems, its long-time behavior
is dominated by the leading isolated eigenvalue $\lambda_1$ --
corresponding to the RP resonance in the TDL -- which is strictly smaller than
$1$, $|\lambda_1| < 1$. That is, $\Mr \approx \lambda_1 \P_1$, where
$\P_1=\frac{\ket{\R_1}\bra{\L_1}}{\braket{\L_1}{\R_1}}$ is the projector to the
corresponding eigenspace, with $\ket{\R_1}$ and $\ket{\L_1}$ being the right and
left eigenvectors. While the concept of RP resonances is rather
old~\cite{Pollicott85,Ruelle86,gaspard,Braun}, and likewise the truncated
propagator has been used before~\cite{Prosen}, fully understanding its utility
in explaining how quantum chaos arises, which is the main focus of this paper,
is new. Importantly, we, for the first time, provide an explicit
quantification of fractal structures responsible for quantum many-body
chaos.

We find that the underlying reason for relaxation is embodied in the properties
of eigenvectors of $\Mr$, \dif{specifically, in how populated basis
operators with support on $s$ consecutive sites are, i.e., $s$ measures how
``many-body'' an operator is}. Denoting by $b_{\ba}$ the components of the
eigenvector $\ket{\R_1}$, we show that its partial norms $w_s:=\sum_{\ba,{\rm
sup}(\ba)=s} |b_{\ba}|^2$, measuring the weight of $s$-body components
(normalization is $\sum_{s=1}^r w_s=1$), grow
with the support size $s$ asymptotically exponentially as $w_s \sim
\mu^s=\e{\dx s}$ \dif{in one-dimensional (1D) systems}. Here, $\dx=\ln{\mu}$ is the \dif{Hilbert space growth
exponent that tells us how much more populated operators with larger
support are, $w_{s+1}/w_s=\mu=\e{\dx}$, and can be interpreted as a many-body fractal dimension}. We note that self-similarity of
eigenvectors has been briefly mentioned in Ref.~\cite{Prosen}. A similar property
also holds for left eigenvectors $\ket{\L_1}$. \dif{In fact, one can capture a
diverging form of both eigenvectors in a single quantity $\kappa_1:=
\frac{\nor{\L_1}\cdot\nor{\R_1}}{|\braket{\L_1}{\R_1}|}$, which is nothing but
the condition number (known in, e.g., numerical linear algebra) of the leading RP
resonance. We find that $\kappa_1$ in 1D systems grows exponentially with the
truncation size $r$, $\kappa_1 \sim \e{\dx r}$, with the same exponent as partial norms, and is
numerically often better behaved than individual eigenvector's partial norms.}

\difa{The condition number $\kappa_1$ is important also because it measures the
strength of the leading eigenprojector $\mathcal P_1$. Since both $\kappa_1
\sim\mu^r$ and $w_s \sim \mu^{s - r}$ (the additional $\mu^{-r}$ comes from
normalization) diverge with the same rate, we argue that for local
observables $A$ with some support $s$, the expansion coefficient $c_1 =
\bracket{A}{\mathcal P_1}{A}\sim w_s \kappa_1 \sim \mu^s$ (recall, $C(t) \approx
c_1 \lambda_1^t$) is finite in the TDL $r \to \infty$. In other words, provided that the
leading eigenvalue is reasonably converged, the RP description gives an accurate
prediction for $C(t)$ of local observables at \textit{arbitrary time}.}

On a practical side, we also find that the convergence of the leading RP
resonance with the support size $r$ is exponential in chaotic systems, similarly
as the convergence of partial binorms $\braket{\L_1}{\R_1}$, making the method
numerically attractive. \dif{This explains the efficiency of the truncated
operator method and also explains why even at modest $r$ it is able to correctly
predict the long-time behavior of correlation functions.}

We also show that unitarity imposes a constraint between $\lambda_1$ and $\dx$, namely
\begin{equation}
 \dx \vv \approx \dt,
\end{equation}
where \dif{$\vv$ is the effective operator causal-cone velocity (the butterfly velocity)} and $\dt=-\ln{|\lambda_1|}$
is the temporal decay rate of dynamical correlations, $C(t) \sim \e{-\dt t}$.
\dif{In exactly solvable cases (dual-unitary circuits in this paper and CMV
matrices Ref.~\cite{CMV}), we find an exact equality and conjecture that the equality
is asymptotically exact at large supports in systems with an isolated leading RP
resonance, with numerical violations at finite $r$ typically being small.}

\dif{Besides 1D quantum circuits without any symmetries, we also study systems
with conserved quantities, for instance total magnetization or energy, and
demonstrate that the above scenario stays valid -- initially local operators with
a finite quasimomentum $k$ again relax due to a fractal operator cascade.}
\dif{We, furthermore, also discuss qubit systems in more than $d=1$ spatial dimension. There
the divergence is milder, $\kappa_1 \sim \e{\dx r^{1/d}}$, where
$r^{1/d}$ measures the linear size of the truncation and $\dx$ is due to
unitarity again related to $\lambda_1$. This prediction is numerically checked
in the 2D kicked Ising model. }

\dif{While we focus on the leading RP resonance, we also find that the spectrum
of $\Mr$ splits into RP resonances that carry physical information, are
independent of truncation details, and have relatively low condition numbers,}
and a bulk composed of eigenvalues that are more badly conditioned and can be
modeled with a random-matrix-like noise stemming from the truncation errors.
While this means that in the $r \to \infty$ limit the non-normal matrix $\Mr$
becomes badly conditioned, due to non-normality being weak, one nevertheless does
not need to worry about the pseudospectrum (a stable generalization of the
spectrum relevant for non-normal operators~\cite{trefethen}), \dif{even at
relatively large truncations up to $r=18$ required in our work.}

\dif{The quantum chaos scenario that we reveal has some similarities with
classical chaotic systems, both dissipative and conservative, but also some
important differences. Up to a contracting prefactor $\lambda_1^t$, the leading
eigenvector $\ket{\R_1}$ with its fractal structure can be considered an
``attractor'' \difa{or an ``unstable manifold''} in the sense that any operator
with a local initial support will with time converge towards $\ket{\R_1}$.
Evolving backward in time, the same can be said for $\ket{\L_1}$. \difa{Importantly,
the fractality is this time not in space but in the many-body support of
operators.} \difa{What is more, in classical systems, one can indeed see that the leading RP eigenvector
$\ket{\R_1}$ (obtained either analytically or through a coarse-graining procedure
bearing similarities to the truncated propagator) localizes on the unstable manifold,
as seen in Ref.~\cite{blum_agam} for the perturbed cat map, in Refs.~\cite{weber1, weber2}
for the kicked top and in Ref.~\cite{gaspard95} for a prototypical example of the pitchfork bifurcation.
See also Ref.~\cite{blank_keller_liverani} for a rigorous treatment of Anosov maps.}

The
fact that $\dx>0$, i.e., $\mu>1$, is a direct embodiment of an effective
``dissipative'' relaxation within perfectly unitary evolution. As in
classical systems, there is no contradiction though. In classical chaotic
systems, relaxation emerges due to the flow from large to small spatial
wavelengths of size $\epsilon$, with the corresponding scaling quantified by the
fractal dimension. Mathematically, this is allowed by an infinite dimensional
functional space describing densities. On the other hand, in our quantum
many-body case the infinite dimensionality is provided by the TDL (diverging
number of qubits) and the role of the hard-to-measure small spatial scale $\epsilon$
is played by a hard-to-measure many-body support size $s$. One has a many-body
Kolmogorov cascade~\cite{kolmogorov}, similar to classical turbulence where the
energy flows from large eddies to increasingly small ones, ultimately resulting
in dissipation~\cite{Richardson}. This time, though, the cascade is not in the
spatial size but rather in the ``many-bodyness'': local operators flow with time to
increasingly non-local ones, accompanied by the scaling relation connecting
their fractal Hilbert space structure (Fig.~\ref{fig:diagram}, right) with the
temporal decay of correlations. Those operators ``escape to infinity'' and asymptotically never return back to the local ones. We stress that this does not violate unitarity
-- while in classical few-degree-of-freedom systems unitarity enforces the
folding mechanism, in the TDL of our qubit systems one can avoid the
``folding'' mechanism as for instance epitomized by the bilateral shift
operator, $U = {\sum_{n = -\infty}^\infty} \ket{n}\bra{n+1}$, which describes
objects coming from and escaping to infinity, and is nevertheless unitary ($UU^\dagger=U^\dagger U=\1$). In this sense the mechanism for
relaxation in quantum many-body systems is quite different from the one in classical chaotic few-particle systems.}

\dif{We note that the picture of local operators evolving to increasingly
non-local ones, thereby leading to chaos and relaxation, is not new, see for
instance Refs.~\cite{tibor22b,curt18,adam18,vedika18,claeys_krylov}. Through the study of properties of RP eigenvectors,
we, however, provide objects that quantify the cascade in closed (unitary)
systems, and are numerically accessible and important for classical simulation of generic many-body systems.}

\section{Operator Kolmogorov cascade}
\label{sec:main}

We would like to describe and understand dynamics, for instance relaxation, in
systems of locally interacting qubits in the TDL, i.e., when the number of qubits $L
\to \infty$. \dif{We will start by discussing 1D circuits without any local
conserved quantities in this section, but will then also explore systems with conservation laws,
e.g., magnetization and energy, as well as systems in more than 1 spatial
dimension in Sec.~\ref{sec:beyond}, showing that the cascade is present also in those systems and is
therefore generic in chaotic systems.}

As a concrete example of 1D systems, we shall demonstrate our theory in quantum
circuits like the kicked Ising model, dual-unitary circuits, and brickwall
circuits with a randomly chosen 2-qubit gate. Such systems are of immediate
interest to modern-day noisy quantum computing
experiments~\cite{googlescrambling21,IBMutility23,IBMint25,feig25,IBMDU26}. The kicked Ising model is one
of the simplest many-body quantum models that displays chaotic behavior for
appropriate parameters (we will use $h_{\rm x}=0.9$, $h_{\rm z}=0.8$, and
different $\tau$, the same as in Ref.~\cite{RP24}). Its propagator for one
time step is
\begin{eqnarray}
  U&=& U_{\rm z} U_{\rm x},\\
  U_{\rm x}&=& \prod_j {\rm e}^{-\ii \tau \hx \sx_j},\nonumber \\
  U_{\rm z}&=& {\rm e}^{-\ii \tau \sum_j \sz_j \sz_{j+1}}\prod_j {\rm e}^{-\ii \tau \hz \sz_j}. \nonumber
  \label{eq:KIU}
\end{eqnarray}
It can be viewed as a circuit composed of a layer of commuting 2-qubit gates (the
zz interaction) and a layer of 1-qubit gates (x and z rotations). Integer
time $t$ counts the number of applications of the Floquet propagator $U$.

We shall work in the Heisenberg picture where self-adjoint observables evolve.
For instance, the operator $A$ at time $t=1$ is
\begin{equation}
  A(t=1)=U^{\dagger} A U = \M A,
\end{equation}
where $U$ is a single time-step unitary state propagator, while $\M$ is a
unitary operator propagator (sometimes called a superoperator). Using the ket-bra notation for elements of the Hilbert space of
operators (and its dual), we can write the
time-evolved operator as $\ket{A(t)}=\M{(t)} \ket{A}$. For simplicity, we focus on
time-independent circuits in which $U(t)=U^t$, and similarly for $\M$. An
infinite temperature expectation value is
\begin{equation}
  \ave{A}=\tr{A}/2^L,
\end{equation}
while a correlation function is ($\ave{A}=0$)
\begin{equation}
  C(t)=\frac{1}{L}\ave{A A(t)},
\end{equation}
where the prefactor ensures that, for extensive $A$, the correlation function is
$L$-independent. In terms of $\M$, it can be simply expressed as
\begin{equation}
  C(t)={\frac{1}{L}}\bracket{A}{\M^t}{A},
  \label{eq:CM}
\end{equation}
with $\braket{A}{B}=\tr{A^\dagger B}/{2^L}$.

As we already argued, revealing chaos and relaxation from the spectral
properties of unitary $\M$ is not easy -- its eigenvectors are complicated
many-body objects. The difficulty is even greater in the TDL, where the spectrum
of $\M$ (for mixing systems) becomes continuous, and its eigenvectors become
non-normalizable, i.e., they cease to exist in the space of normalizable
vectors $\l2 = \{\psi; \sum_j \abs{\psi_j}^2 < \infty\}$. Formally, one can
then write the spectral decomposition using spectral measures of projectors $\mathcal
P(\vartheta)$, $\M = \int_{0}^{2\pi} e^{\ii \vartheta} \mathrm{d}\mathcal
P(\vartheta)$~\cite{functional}, but they are cumbersome to use in practice.

Following a more physical intuition, on the level of eigenvalues of $\M$, exponential
relaxation $C(t) \sim \lambda^t$ for large $t$, with $\lambda<1$, would by
Eq.~(\ref{eq:CM}) suggest some subunitary eigenvalue~\cite{foot_unit}. However, a simple
argument shows that if $\M$ had such an eigenvalue $|\lambda|<1$, it
cannot be from $\l2$. That is, if we find an eigenvalue $\lambda$ and the
associated eigenvector $\ket{\psi}$, $\M\ket{\psi}=\lambda \ket{\psi}$, then for
unitary $\M$ on $\l2$, $\M^{-1} \M=\1_{\l2}$, one has $\bracket{\psi}{\M^\dagger
\M}{\psi}=|\lambda|^2 \braket{\psi}{\psi}=\braket{\psi}{\psi}$. Therefore,
provided $|\lambda|^2 \neq 1$, the norm $\braket{\psi}{\psi}$ cannot be finite
and such an ``eigenvector'' $\ket{\psi}$ cannot be from $\l2$.

The contradiction above can be circumvented by allowing $\braket{\psi}{\psi}$ to
diverge. This can be rigorously formalized with the language of rigged Hilbert
spaces~\cite{gelfand,ballentine,bohm,demadrid_rigged}, where we essentially extend the considered Hilbert
space ${\cal H}$ with generalized vectors with diverging norms (forming a space of linear functionals ${\cal S}^\times$), but well-defined inner products with a chosen subset of vectors (the observables ${\cal S}$), ${\cal S} \subset {\cal H} \subset {\cal S}^\times$. The most well-known example of this
are generalized functions, also called distributions (e.g., the Dirac delta and
its derivatives), that cannot be normalized, but have well-defined inner products
with smooth enough functions.

In this language, the exponential decay of correlation functions is then
actually controlled by subunitary ``generalized eigenvectors'' of $\M \approx \sum_j
\lambda_j \frac{\ket{\R_j}\bra{\L_j}}{\braket{\L_j}{\R_j}}$, $\abs{\lambda_j} <
1$. We refer to $\lambda_j$ as RP resonances, while the generalized eigenvectors
$\R_j,\L_j \in {\cal S}^\times$ are sometimes called right and left Gamow
vectors (especially in the context of scattering)~\cite{demadrid_gamow}. Gamow
vectors and their singular properties have been studied before in explicitly
solvable classical systems~\cite{Hasegawa92,Tasaki93,gaspard95,
blank_keller_liverani,antoniou97}. On the quantum side, no explicit solutions
are known \difa{apart from CMV matrices inspired by this paper~\cite{CMV}}. However, a fruitful numerical procedure of extracting RP resonances
is choosing ${\cal S}$ to be local observables and constructing the so-called
truncated propagator by projecting $\M$ to ${\cal S}$~\cite{Prosen}, the intuition
being that relaxation takes place only for local observables.

\subsection{The truncated propagator}
\label{sec:Trunc}

Considering translationally invariant lattice systems whose propagator $\M$
commutes with a one-site translation superoperator ${\cal T}$, we can label
operator basis elements by their continuum quasimomentum index $k\in [-\pi,\pi]$
that labels eigenvalues $\e{\ii k}$ of ${\cal T}$ (limiting, for simplicity, to 1D systems). Basis operators $\Aa{\alpha}$
are indexed by the Pauli string index $\ba=(\alpha_1,\ldots,\alpha_r)$ that encodes
the operator's density $\a{\alpha}$,
\begin{equation}
  \Aa{\alpha}=\sum_j \e{\ii k j} \a{\alpha}_j,\quad \a{\alpha}_j=\sigma_j^{\alpha_1}\cdots \sigma_{j+r-1}^{\alpha_r},
  \label{eq:Aa}
\end{equation}
where the sum over $j$ runs over all sites in an infinite lattice, and
$\sigma^{\alpha_j}\in \{\1,\sx,\sy,\sz \}$ are Pauli matrices spanning the local
operator basis. Note that the operators $\a{\alpha}_j$ in Eq.~\eqref{eq:Aa} are
simple translations of the same site-independent density; e.g., for $k=0$ and
$\a{\alpha}_1=\sz_1$ we would get the total magnetization $A=\sum_j \sz_j$. The
matrix elements of $\M$ are in turn
\begin{equation}
  [\M]_{\ba,\bb}=\bracket{\ba}{\M}{\bb},
  \label{eq:Mab}
\end{equation}
where we suppressed an implicit dependence of $\M$ on the (quasi)momentum $k$, and $\ket{\ba}=\ket{\Aa{\alpha}}$.

Truncation to the basis of locally supported operators now means that we only
take into account Pauli strings $\ba$ with at most $r$ indices. Furthermore, we
can exploit translational invariance to simplify the expression for $[\M]_{\ba,\bb}$.
Namely, because local densities $\a{\alpha}_j$ are propagated in the same
way at all positions $j$, we have
\begin{equation}
[\Mr]_{\ba,\bb}=\braket{\sum_{p=-v}^v \e{\ii k p} \a{\alpha}_p}{U^\dagger \a{\beta}_0U},
\label{eq:Mr}
\end{equation}
where the Pauli strings are limited to those with support on at most $r$ sites
in order to get our truncated propagator $\Mr$. We can see that one needs to
propagate only $\a{\beta}_0$ that starts at site $0$ and then calculate its
overlap with all shifted operators, where the maximal number of shifts $v$
depends on the circuit in question. It is equal to the maximal number of sites
that the operator can spread to the left, or to the right, of its original
support, under the one-step propagator $\M$. This light-cone velocity $v$ is
$v=1$ for the kicked Ising model and $v=2$ for brickwall circuits. For a finite
maximal support size $r$, the truncated propagator $\Mr$ is a finite square
matrix of dimension $N=3\cdot 4^{r-1}$. That is, because the first Pauli index
$\alpha_1$ marks the beginning of the local density, it should not be an identity $\1$ and can therefore take $3$ possible values, whereas all other
$\alpha_{j>1}$ can take 4 possible values. This means that our basis does not include the identity, and, therefore, the corresponding trivial eigenvalue $1$ is not in the spectrum of $\Mr$.

To get the matrix elements of $\Mr$ in an infinite system, one therefore needs to
do evolution of local operators only on $r+2v$ sites (with some tricks, which we use on 1D and 2D kicked Ising models, evolution on only $r$ sites suffices~\cite{prx}). For further technical details about the construction of
$\Mr$ see Refs.~\cite{RP24,urban}. Importantly, $\Mr$ in
Eq.~(\ref{eq:Mr}) is written exactly in the TDL, i.e., there is no finite $L$
involved, apart from the truncation to operators with their density supported on at
most $r$ sites. While its size grows exponentially with $r$, \dif{using standard
techniques for finding the largest eigenvalue of a sparse matrix (Arnoldi iteration based,
or simply the power method)} we will be able to numerically study truncations up
to $r=14\text{--}18$ which will be enough to see the true TDL behavior of local
operators in chaotic many-body quantum systems. \dif{Observe also that because
we get the spectrum for all quasimomenta $k$, i.e., we cover the full operator
basis, we can in principle treat dynamics of any operator. For instance, a
strictly local operator would be a superposition (Fourier integral) of all
$k$-modes.}

The truncated propagator has been used before: first for $k=0$ in
Ref.~\cite{Prosen}, and more recently in a cellular automaton in
Ref.~\cite{katja}. Including its quasimomentum extension~\cite{RP24}, it is a
handy tool for studying chaotic as well as integrable models, for instance, finding new
integrable circuits~\cite{U1}, inhomogeneous conserved operators in an otherwise
homogeneous system~\cite{kpzwall}, efficiently extracting the diffusion
constant~\cite{urban,prx}, studying long timescales possibly involved in the
breakdown of conservation laws~\cite{prx}, \dif{relaxation to equilibrium~\cite{hamazaki}}, and classifying different dynamical
properties of cellular automata~\cite{rustem}. It can be viewed as a particular
scheme to extract Ruelle-Pollicott
resonances~\cite{Pollicott85,Ruelle86,gaspard}. An alternative method to do the
same in a many-body context is via the weak-dissipation limit~\cite{Mori}, see also
Refs.~\cite{sarang,curt,lucas,diego} (and for a different approach
Ref.~\cite{lychkovskiy}), which, though, is, in our experience~\cite{urban},
numerically less efficient and versatile than the truncated propagator approach.
It also remains to be seen how the formalism
can be carried over to multi-point correlation functions, see Ref.~\cite{felix}
for some recent results.

\difa{The truncated propagator formalism has some similarities to the
Krylov-based methods~\cite{krylov}; however, there are also significant
differences. While the Krylov space method can be understood as a Taylor
expansion for a specific operator that is valid for short enough times, we
treat the whole operator space in one stroke and, importantly, provided the RP
resonances converge (when and why they do is one of the main results of this
paper), we are not limited to short times.}


\subsection{Long-time behavior and the fractal cascade}

Because the truncated propagator $\Mr$ is a projection of the unitary $\M$, its
spectrum lies within the unit circle. Let us denote its eigenvalues by $\lambda_j$
and the corresponding right and left eigenvectors by $\ket{\R_j}$ and $\ket{\L_j}$
(for finite $r$, it is generically diagonalizable despite being a non-normal
matrix). We can then write its spectral decomposition as
\begin{equation}
  \Mr=\sum_j \lambda_j \P_j,\qquad \P_j=\frac{\ket{\R_j}\bra{\L_j}}{\braket{\L_j}{\R_j}},
\end{equation}
where $\P_j$ are the eigenprojectors~\cite{footP}. Note that while biorthogonality
$\braket{\L_j}{\R_p} \propto \delta_{j,p}$ holds, there is some freedom in the
choice of normalization, $\nor{\R_j}^2:=\braket{\R_j}{\R_j}$, and we write
$\P_j$ in the normalization-independent form (one can enforce two conditions,
common choices are $\braket{\L_j}{\R_j}=1$ and $\nor{\R_j}=1$, or $\nor{\R_j}=1$
and $\nor{\L_j}=1$ while $\braket{\L_j}{\R_j}$ is ``free''). The autocorrelation
function of an extensive observable $A$ defined in terms of a single Pauli
string $\ba$ is now simply
\begin{equation}
  C(t)=\bracket{\ba}{\Mr^t}{\ba}=\sum_j c_j \lambda_j^t,
  \label{eq:CMr}
\end{equation}
where the expansion coefficient $c_j$
\begin{equation}
  c_j=\frac{\braket{\ba}{\R_j}\braket{\L_j}{\ba}}{\braket{\L_j}{\R_j}},
  \label{eq:cj}
\end{equation}
is manifestly independent of the chosen normalization. The approximation with
$\Mr$ (\ref{eq:CMr}) will be good provided the truncation size $r$ is large
enough.

We are interested in chaotic systems in which autocorrelation functions decay
exponentially, which is, in our formalism with $\Mr$, reflected in the fact that the
largest eigenvalue (in modulus) $\lambda_1$ -- the so-called leading RP
resonance -- is isolated from the rest, $1>|\lambda_1|> |\lambda_{j\ge 2}|$. Consequently, the long-time decay will be
\begin{equation}
  C(t) \approx c_1 \lambda_1^t \sim \e{-\dt t},\qquad \Mr^t \approx \lambda_1^t \P_1,
  \label{eq:dt}
\end{equation}
where we have defined the temporal decay rate $\dt := -\ln{|\lambda_1|}$.
If we apply Eq.~\eqref{eq:dt} at the level of operators, we obtain $\Mr^t
\ket{\ba} \approx \lambda_1^t c_1 \ket{\R_1}$. One can, therefore, interpret $\ket{\R_1}$
as a positive-time ``attractor'', to which all local operators converge after a long enough time
(before decaying completely due to the $\lambda_1^t$ prefactor).

Note that there is an important subtle point implicitly used in the above
argument leading to keeping only $\P_1$. Namely, $\Mr$ is a non-normal operator
for which it can happen that the decay in the thermodynamic limit is {\em not
given} by the leading $\lambda_1$. This can occur because the expansion
coefficients can grow exponentially large with increasing $r$, invalidating the
approximation with only $\lambda_1$ in the correct TDL when one first takes $r
\to \infty$ and only then $t \to \infty$. In such situations, it might be better
to look at the pseudospectrum~\cite{trefethen} rather than at the spectrum. We
will return to this point in Sec.~\ref{sec:kappa}. For now let us just say that
in our truncated propagator, one does not need to worry about the
pseudospectrum.

Since classical chaos is defined (in any of its different ways) in terms of
long-time properties $t \to \infty$, if there is any analogous ``chaos'' in
our many-body quantum circuits, it should be embodied in the properties of
$\P_1$, that is, the right and left leading eigenvectors $\ket{\R_1}$ and
$\ket{\L_1}$, and the corresponding eigenvalue $\lambda_1$. In the following, we
shall therefore focus on the properties of those, revealing the underlying
fractal structures that lead to relaxation and chaos.

\begin{figure*}[t!]
  \centerline{\includegraphics[width=0.99\textwidth]{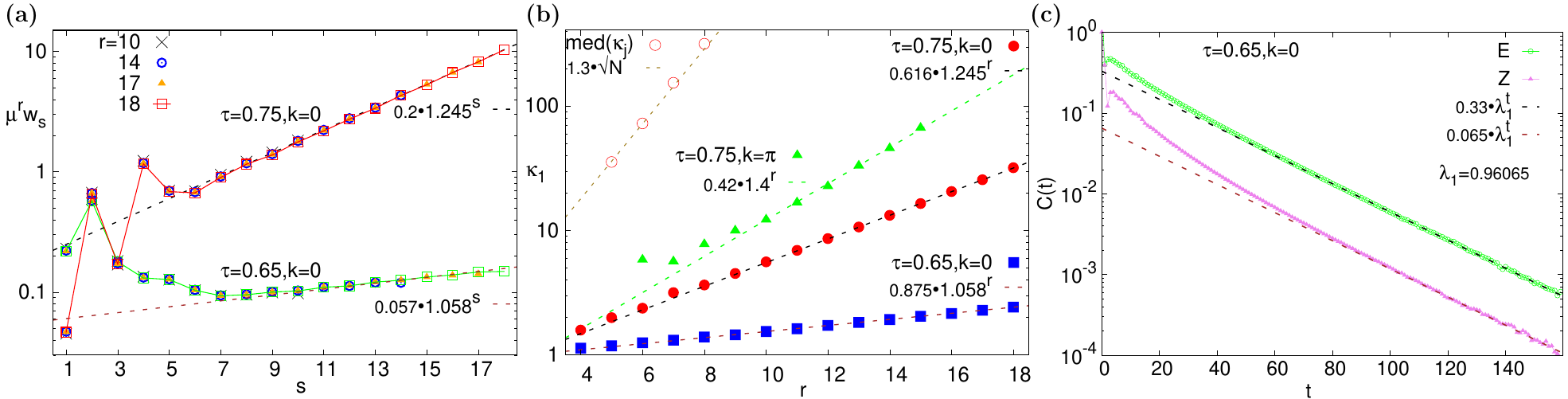}}
  \caption{{\bf Fractal cascade in the 1D kicked Ising model}. (a) Partial
  norms $w_s$ (\ref{eq:ws}) of the normalized leading right
  eigenvector $\ket{\R_1}$ of the truncated propagator $\Mr$ diverge as $\sim
  \mu^s$ with the operator density support size $s$. On the vertical axis, the
  same values of $\mu$ are used for scaling as for the dashed lines. (b) The leading
  condition number $\kappa_1$ (\ref{eq:kappaj}) for different $\tau$ and
  quasimomenta $k$ (full symbols), all diverging with the truncation parameter
  $r$ as $\kappa_1 \sim \mu^r$ (\ref{eq:kappa1}) with the same $\mu$ as used in
  (a). Empty red circles are the median $\kappa_j$ (for $\tau=0.75$, $k=0$),
  showing a clear separation from $\kappa_1$ (full red circles), further
  discussed in Sec.~\ref{sec:kappa}. (c) \dif{Prefactor $c_1$ in the asymptotic decay of autocorrelation functions is different for an observable with $s=2$ (green) and $s=1$ (violet); data taken from Ref.~\cite{RP24}.} All plots are for parameters $\hx=0.9,\hz=0.8$.}
\label{fig1}
\end{figure*}

\subsubsection{Diverging leading eigenvector}
\label{sec:div_vec}

Because we have $|\lambda_1|<1$, i.e., $\dt>0$, we know that in the $r \to \infty$
limit, $\ket{\R_1}$ (and $\ket{\L_1}$) cannot be from $\l2$. Remember that
$\ket{\R_1}$ is an operator towards which any local operator $A$ with a finite
overlap $\braket{\L_1}{\ba}$ will evolve forward in time, $t \to \infty$,
while $\ket{\L_1}$ is a similar asymptotic operator for evolution backward in
time, $t \to -\infty$. We are going to show that this non-$\l2$ divergence is
specifically exponential in systems with decaying correlations (say chaotic) and
that the rate has a clear physical significance.

In order to understand properties of $\ket{\R_1}$, we study its partial norms $w_s$,
corresponding to components in $\ket{\R_1}$ with support on $s$ sites, defined as
\begin{equation}
  w_s:=\sum_{\ba,{\rm sup}(\ba)=s} |b_{\ba}|^2,\qquad b_{\ba}=\braket{\ba}{\R_1},
  \label{eq:ws}
\end{equation}
where $b_{\ba}$ are expansion coefficients of $\ket{\R_1}$ in the local operator
basis, and ${\rm sup}(\ba)$ is the support size of the $\ba$-th basis element
(the number of basis elements with support $s$ is $N_s=3^2\cdot 4^{s-2}$ for
$s>1$ and $N_1=3$ for $s=1$; $\sum_{s=1}^r N_s=N$)~\cite{foots}. Using normalized
$\nor{\R_1}=1$, i.e., $\sum_{s=1}^r w_s=1$, for the eigenvector to be in $\l2$
one would need $w_s$ to decay sufficiently fast with increasing $s$. However, in
our case, this does not happen -- instead, partial norms {\em grow exponentially}
as $w_s \sim \mu^s$ with the support $s$ of operator density\dif{~\cite{footws}}. This can be seen
in Fig.~\ref{fig1}(a), where we show results for the kicked Ising model in the
chaotic regime. What is more, for a sufficiently large $r$
(for the parameters shown, $r=10$ is already enough on the scale of the plot, while
$r=6$ would not yet be so) there is a simple scaling form for large $s$ in terms
of both $s$ and $r$ (a prefactor left out is fixed by the normalization and is model
dependent, but independent of $r$ and $s$)
\begin{equation}
  w_s \approx \mu^{s-r},
  \label{eq:expws}
\end{equation}
with $\mu>1$. The very same scaling form, i.e., with the same $\mu$, is obtained
also for the left eigenvector $\ket{\L_1}$ (data not shown).

Looking at the expansion coefficient $c_1$ in Eq.~(\ref{eq:cj}), we see that in
the numerator, we have a term that will scale similarly to $|b_{\ba}|^2$
(forgetting for a moment that $\R_1$ and $\L_1$ are different), i.e., as $w_s$,
while we have the binorm $\braket{\L_j}{\R_j}$ in the denominator. In fact, the condition number
$\kappa_j$ of the eigenvalue $\lambda_j$ has a similar form
\begin{equation}
  \kappa_j := \frac{\nor{\R_j}\!\cdot\!\nor{\L_j}}{|\braket{\L_j}{\R_j}|}=\sqrt{\tr{\P_j \P_j^\dagger}}.
  \label{eq:kappaj}
\end{equation}
The condition number, implicitly depending on $r$, measures the size of the
projector (by the Hilbert-Schmidt, i.e., the Frobenius norm). For unitary or
Hermitian matrices, where the left and right eigenvectors are the same, one has
$\kappa_j=1$, while for non-normal matrices, like our $\Mr$, it is larger, $\kappa_j
\ge 1$. If we use normalization $\nor{\R_j}=1$ and $\nor{\L_j}=1$, it is equal to
the reciprocal binorm, $\kappa_j=1/|\braket{\L_j}{\R_j}|$. The condition number,
known in numerical linear algebra~\cite{watkins}, tells us how sensitive
eigenvalues are to perturbations~\cite{stewart}. For a sufficiently small
perturbation $V$ of a matrix $M$, the change in the eigenvalue $\lambda_j$ of $M$
is upper bounded~\cite{foot2} by $|\Delta \lambda_j| \le \kappa_j \nor{V}_2$.
Considering its relevance for perturbations, it has found its use in various
fields of physics where non-normality appears, for instance, dissipative
resonators and lasers~\cite{petermann,schomerus}, or in scattering
problems~\cite{savin}.

Putting the numerator and denominator in Eq.~(\ref{eq:cj}) together and assuming
that all expansion coefficients with a given $s$ are approximately equal, we can argue
that $c_1$ will scale as $c_1 \sim w_s\kappa_1$ (with $s$ corresponding to
the chosen $\ba$). Therefore, in order for $c_1$ to be finite in the TDL, in
turn causing correlations to decay with $\lambda_1$ (rate $\dt$), we need a
nonzero $\lim_{r \to \infty} w_s \kappa_1$. In Fig.~\ref{fig1}(b), we look
at the scaling of $\kappa_1$ with $r$ in the same kicked Ising model, finding
that in all chaotic cases, we have exponential divergence with the same $\mu$ as
for $w_s$, that is
\begin{equation}
  \kappa_1 \sim \mu^r.
  \label{eq:kappa1}
\end{equation}
Because the scaling we used on the $y$ axis in Fig.~\ref{fig1}(a) in order to
achieve a collapse of all the data has the same $\mu$ as in the divergence of
$\kappa_1$ in Fig.~\ref{fig1}(b), this means that $c_1$ for local
observables is finite in the TDL. In other words, to have
an exponential decay of correlations with the rate $\dt$ (\ref{eq:dt}), one needs
diverging partial norms $w_s \sim \mu^s$ of the corresponding left and right
eigenvectors and a diverging condition number $\kappa_1 \sim \mu^r$ with
the same $\mu$.

At small $s$, where we are not yet in the asymptotic exponential regime, one can
see in Fig.~\ref{fig1}(a) a rather different behavior that depends on the system's
parameters, e.g., $\tau$. Taking $\tau=0.65$ and $k=0$ as an example, we
notice that $w_{s=2}$ is considerably larger than $w_{s=1}$. This is in turn
reflected in a larger coefficient $c_1$ in Eq.~(\ref{eq:dt}) for a typical
observable with support on 2 sites compared to 1-site observables. Physical
consequences of this are
illustrated in Fig.~\ref{fig1}(c). We see that the asymptotic decay is perfectly
in-line with the theory in Eq.~(\ref{eq:dt}) without any fitting parameters --
for the dashed lines we use the calculated $\lambda_1$ (see Fig.~\ref{fig:fig2b}),
while $c_1$ is obtained from the overlap with the leading eigenvector
$\L_1$. For details on the observables used, see Ref.~\cite{RP24} from where we
take the data. We also see that for the $s=1$ observable, where $c_1$ is
smaller, the asymptotic decay with $\lambda_1$ starts at larger times. This
means that $w_s$ in a plot like Fig.~\ref{fig1}(a) already gives us hints for
which observables it will be harder to see the asymptotic exponential decay. An
extreme case is very small $w_{s=1}$ for $\tau=0.75, k=0$, which is indeed
what was observed~\cite{RP24}.

The singular behavior of $\R_1$ (and $\L_1$) means that the operator towards
which any initial operator with a local density converges, i.e., to which it is attracted
as $t \to \infty$, is increasingly a many-body object -- the probability weight
of components grows with increasing support size $s$ (e.g., taking $\R_1$ at a
given large truncation $r$). \dif{As we shall discuss, this exponential growth
with $\mu>1$ is on one hand an indicator of chaos, but on the other hand also
means that using the truncated propagator $\Mr$ to evolve observables,
$\ket{A(t)}\approx \Mr^t \ket{A(0)} \sim \lambda_1^t \ket{\R_1}$, is not a good
approximation of the exact $A(t)$ in terms of norms -- at large $t$ we are
losing all operators with density supported on more than $r$ sites -- but is
rather an accurate description if projected back to local observables $\braket{A(0)}{A(t)} = C(t)$ (see also Ref.~\cite{tibor22b} for a random-circuit-based
study with similar conclusions). In fact, this is what one can generically
expect for Pauli-string-based truncation methods in deterministic systems for
large times, provided they are implemented appropriately, whereas error bounds
for the entire $\ket{A(t)}$ can be proven only for finite times~\cite{xu} and in
the presence of noise~\cite{shao, yao,pauli_prop_npj} or
randomness~\cite{pauli_prop_prl}.

  \begin{figure}[t!]
    \centerline{\includegraphics[width=3.0in]{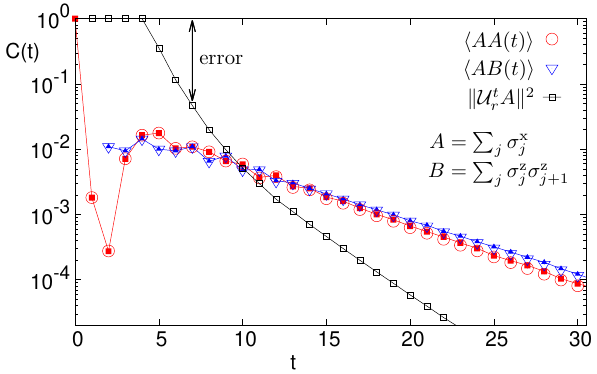}}
    \caption{\dif{\textbf{Correlation functions with the truncated propagator.} Small full red squares and full blue triangles are exact $C(t)$ in the TDL, whereas large empty red circles and blue triangles are results obtained by $\Mr$ with $r=10$. Small black squares show that most of the norm is lost after $t=4$ when the evolved $\sigma_j^{\rm x}(t)$ starts being supported on more than $10$ sites. 1D kicked Ising model at $\tau=0.75$.}}
    \label{fig:2dcorr}
\end{figure}
Let us numerically demonstrate that in our implementation of the
truncated propagator, where we work in the TDL and there are no finite size
effects, description with $\Mr$ can be accurate at all times even for modest
truncations $r$, despite losing most of the norm in $\ket{A(t)} \approx \Mr^t
\ket{A(0)}$. In Fig.~\ref{fig:2dcorr}, we show data for a deliberately small
truncation $r=10$, comparing exact correlation functions (computed for a random state of $L=35$ qubits for which there are no finite-size effects for the times shown, see Ref.~\cite{RP24} for details) with the ones obtained
with the truncated propagator as $C(t)\approx \bracket{A}{\Mr^t}{A}$. We can see that the autocorrelation function as
well as the cross-correlation function agree perfectly. The reason for good
agreement at long times already with $r=10$ is that the operator dynamics at
scales larger than $r=10$ is already in the asymptotic chaotic regime where one
has a shift-like dynamics (discussed later) in which the operators are pushed to
larger supports, from which the probability to return back to low supports is exponentially small.
An efficient description with $\Mr$ is
especially important when the decay is slow, for instance in systems with
symmetries, where the truncated propagator can be used to describe the long-time
behavior of correlation functions and obtain transport coefficients with higher
accuracy~\cite{urban,prx} compared to what is possible with other methods.
}

Let us now return back to the fractal structure of leading eigenvectors.
As we shall see in the following subsection, the growth exponent $\mu$ can be
considered as a fractal dimension of the ``attractor'' $\P_1$.

\subsubsection{Fractal dimension}
\label{sec:fractal}

Let us recall the characterization of fractals in $d$-dimensional space by one of the simplest measures,
the box-counting dimension. Taking a box of linear size $\epsilon$, one
measures the number of $d$-dimensional boxes $N(\epsilon)$ covered by a set (a fractal).
The fractal box-counting dimension $d_{\rm B}$ is then defined by the asymptotic scaling $N(\epsilon)
\sim (1/\epsilon)^{d_{\rm B}}$, formally $d_{\rm B}=\lim_{\epsilon \to 0}
\ln{N(\epsilon)}/\ln{(1/\epsilon)}$. \dif{It quantifies the number of occupied boxes in terms of their maximal number $N_{\rm max}=(1/\epsilon)^d$ achieved for $d$-dimensional sets, $N(\epsilon) \sim N_{\rm max}^{d_{\rm B}/d}$.} Structures in classical chaotic systems,
such as a chaotic attractor, or stable and unstable manifolds, have a nontrivial
fractal dimension, resulting in complicated dynamics in which ``information'' flows to finer and finer {\em spatial scales}.

\dif{In quantum many-body lattice systems, the spatial resolution is finite (due
to a finite $\hbar$ and lattice spacing);} however, one can argue that it is
hard to measure many-body observables, that is, observables whose density is
supported on many sites. Therefore, a quantity analogous to the linear size
$1/\epsilon=:L$ in classical systems, which measures the maximal number of
possible boxes $L$ along one dimension, \dif{is the linear support size $s$. In
the TDL of $s \to \infty$, the dimension of the Hilbert space diverges, which is a
prerequisite for any discussion of chaos. The total operator Hilbert space
dimension $N_{\rm max}$ is $4^s$ for a system of $s$ qubits, and the fractal
dimension measures how much of this total space is occupied by a given set
of quantum operators. More explicitly, up to normalization, the number of occupied
operators $N(s)$ with support $s$ is equal to the weight, $N(s) \sim w_s$, which
by the definition of fractal dimension grows as $N(s) \sim N_{\rm
max}^{\dx/\ln{4}}=\e{s\, \dx}$ ($\ln(4)$ in $N_{\rm max}^{\dx/\ln{4}}$ avoids
having to use a base-4 logarithm to get $\dx$). Using Eq.~\eqref{eq:expws} for
$w_s$, we get the fractal many-body dimension $\dx$,
\begin{equation}
  \dx := \lim_{s \to \infty}\frac{\ln{N(s)}}{s}=\ln{\mu}.
\end{equation}
Note a subtle difference between classical fractals in configuration space,
where $N(\epsilon)\sim L^{d_{\rm B}}$ scales polynomially in the linear size
$L$, and our many-body fractality, where one has $N(s) \sim \e{s\,\dx}$, i.e., $N(s)$ is
exponential in the linear support size $s$. The difference comes because the
configuration space is a direct sum of individual 1D spaces and therefore
$N_{\rm max}=L^d$, whereas in the quantum case one has a tensorial product
structure of individual sites resulting in $N_{\rm max}=4^s$, i.e., the
dimension is exponential in the ``linear size'' $s$. Such a definition in terms of
scaling with the total Hilbert space size is, in fact, not new. It is often used
in studies of single-particle (non-interacting) systems, a well-known example
are critical disordered systems~\cite{anderson}, where one often uses the
inverse participation ratio of pure states to measure $N$. It is, however, also
used in interacting many-body systems, see, e.g., Refs.~\cite{bogomolny12,anto13}.
}

\dif{While formally our usage is similar, we importantly use it on the space of
observables and not on pure states. This is in line with our desire to avoid
unmeasurable many-body eigenstates and rather focus on observable quantities.
The asymptotic form of $A(t)$, or even better of correlation functions, does
depend on $w_s$. Namely,} the sizes of $w_s$ are reflected in the prefactors
$c_1$ in front of the
asymptotic exponential decay of correlation functions (recall the discussion
around Fig.~\ref{fig1}(c)), so they can be measured. This does not mean that it
would be easy to experimentally determine $\mu$ in $w_s \sim \mu^s$. In
particular, we see that one will typically have the exponential dependence only
for sufficiently large $s$, e.g., in Fig.~\ref{fig1}(a) for $s \ge 6$ (for
$\tau=0.65, k=\pi$, not shown, a bit earlier, $s \ge 4$). Therefore, to truly
characterize many-body quantum chaos, one would need to be able to measure
$s$-body observables with $s$ larger than $s=1$ or $2$ usually measured.
However, that is expected; true relaxation (and chaos) is anyway expected to manifest
itself only at sufficiently large times, i.e., at sufficiently large spatial
support $s$.

We have defined the fractal dimension $\dx$ in terms of the probability norm,
i.e., squares $|b_{\ba}|^2$. One could also look at other moments,
$|b_{\ba}|^m$, and define a whole spectrum of dimensions $\dx(m)$. We have
checked a particular non-critical set of parameters in the kicked Ising model and do not find
any multifractality, that is, $\dx(m)$ is a linear function of $m$, see
Appendix~\ref{app:KI_mult}.

\subsubsection{Equality of temporal decay and many-body fractality}
\label{sec:fractal_eq}

\difa{We are going to show that $\dx$ is not an independent quantity but is connected to $\lambda_1$
due to unitarity. A non-rigorous argument starts by noticing
that unitarity implies that $\nor{A(t)}$ is independent of $t$. Resolving the norm by
support (akin to our definition of partial norms \eqref{eq:ws}), we write
$\nor{A(t)}^2=\sum_{s = 1}^\infty\sum_{\ba,{\rm sup}(\ba)=s}
\abs{\bracket{\ba}{\M^t}{A}}^2$. Due to the causal cone, the first sum
can be truncated at finite support, i.e., $A(t)$ with support $r_0$ at $t = 0$
spreads \mz{on average} to $r = 2\vv t + r_0$ sites at time $t$, \mz{where $\vv \leq v$ is the effective operator causal-cone velocity, often called the butterfly velocity~\cite{susskind15,curt18,adam18,vedika18}}. Provided that $t$ is
large enough, we can now approximate the matrix elements with only the leading
RP resonance $\bracket{\ba}{\M^t}{A} \approx \lambda_1^t
\braket{\ba}{\R_1}\braket{\L_1}{A}/\braket{\L_1}{\R_1}$, allowing us to express $\nor{A(t)}$ with
partial norms of the leading RP eigenvector
\begin{align}
    \nor{A(t)}^2 \approx \abs{\lambda_1}^{2t}\frac{\abs{\braket{\L_1}{A}}^2}{\abs{\braket{\L_1}{\R_1}}^2} \sum_{s = 1}^{2\vv t + r_0} w_s. \label{eq:norm_a_t}
\end{align}
Taking a ratio of norms at two consecutive times and using the scaling
of partial norms $w_s \sim \mu^s$, we obtain $1=\frac{\nor{A(t+1)}^2}{\nor{A(t)}^2}\approx|\lambda_1|^{2} \frac{\sum_{s=1}^{2\vv(t+1)+r_0}w_s}{\sum_{s=1}^{2\vv t+r_0}w_s}\approx|\lambda_1|^2 \mu^{2\vv}$,}
\difa{leading to}
\begin{equation}
  |\lambda_1|\mu^\vv \approx 1,\quad\hbox{or equivalently}\quad   \dx \vv \approx  \dt.
  \label{eq:ineq}
\end{equation}
The above equality connects the temporal decay of correlations
$\lambda_1$ and the many-body locality scaling factor $\mu$: a fast decay of correlations is reflected in a fast growth of partial norms, i.e., at large
times, the decay of correlations and thereby relaxation is enabled by the
correspondingly fast evolution of local to non-local operators as described by
$w_s \sim \mu^s$.
\dif{We can show analytically that Eq.~\eqref{eq:ineq} is exact in dual-unitary
circuits (see Sec.~\ref{sec:du}), as well as in some solvable CMV
matrices~\cite{CMV} describing asymptotic operator dynamics. In cases where we
have to rely on numerics and finite $r$, we often find a small violation (the
reason for using $\approx$ in Eq.~(\ref{eq:ineq})). This could be understood as
being due to our approximation of taking into account only the leading projector
$\P_1$.
Provided that $\lambda_1$ is isolated, meaning that $|\lambda_1|-|\lambda_2|>0$,
we conjecture that the equality is asymptotic, i.e., it is exact in the limit $r
\to \infty$ \mz{(provided one uses the correct asymptotic $\vv$)}.}

Numerically checking the validity of Eq.~\eqref{eq:ineq} in the kicked Ising
model \mz{where we estimate $\vv \approx 0.9$ at $\tau=0.75$ and $\vv \approx
0.75$ at $\tau=0.65$ (estimates of $\vv$ are obtained from $w_s$ for the evolved $\sigma^{\rm x}(t=7)$)}, we obtain that the product $|\lambda_1|\mu^\vv$ takes
values $1.005$ (at $\tau=0.75, k=0$), $1.02$ (at $\tau=0.75, k=\pi$, data not
shown), $1.002$ (at $\tau=0.65, k=0$), and $1.06$ (at $\tau=0.65, k=\pi$, data
not shown). We observe similar numbers also for generic brickwall quantum
circuits that have light-cone velocity $v=2$ in Appendix~\ref{app:circs}.
In integrable systems and for $\R_j$ corresponding to a conserved operator (from
$\l2$), one has a trivial equality because $\lambda_1=1$ and $\mu=1$, resulting
in $\dt=\dx=0$.

The fractal many-body dimension $\dx$, or, equivalently, the scaling factor $\mu$, is therefore given by the temporal decay rate of correlations $\dt$, Eq.~(\ref{eq:ineq}), similarly to the connection between the Lyapunov dimension (given in terms of positive Lyapunov exponents) and the fractal dimension of classical chaotic attractors~\cite{Yorke,Farmer}. \dif{The diverging nature of the leading RP eigenvectors in local 1D systems can therefore be summarized as
  \begin{equation}
    w_r \sim \kappa_1 \sim \left(\frac{1}{|\lambda_1|^2}\right)^{\frac{r}{2\vv}}.
    \label{eq:kappalam}
  \end{equation}
}

\begin{figure}[t!]
    \centerline{\includegraphics[width=\linewidth]{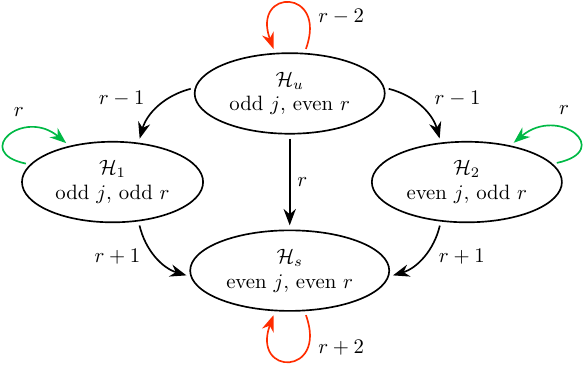}}
    \caption{\textbf{Allowed transitions in dual-unitary
    circuits.} Bubbles denote distinct subspaces of operators with a given parity
    of the first site $j$ they act on nontrivially and the parity of support
    $r$. Arrows represent the possible transitions between operator subspaces
    after applying \textit{one layer} (half time step) of the dual-unitary circuit and aligning
    to the causal cone, $\M_{1/2}\ket{A} := \mathcal{T}\mathcal{U}_\mathrm{odd}\ket{A}$. That is, an arrow pointing from subspace $\mathcal
    H_\text{from}$ to subspace $\mathcal H_\text{to}$ means that
    $\bracket{B}{\mathcal U_{1/2}}{A} \neq 0$ for $A \in \mathcal H_\text{from},
    B \in \mathcal H_\text{to}$ is possible in dual-unitary circuits. The
    labels on the arrows denote the changes of support in the
    transition. The graph is valid for large $r > 4$ and is reproduced based on
    Ref.~\cite{holden_dye}.}
    \label{fig:du_graph}
\end{figure}

\subsection{Dual-unitary circuits}
\label{sec:du}

\begin{figure*}
    \centerline{\includegraphics[width=\textwidth]{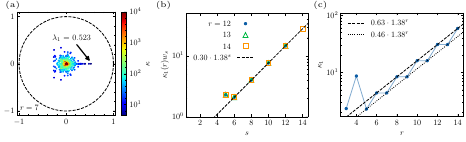}}
    \caption{{\bf Fractal cascade in dual-unitary circuits}. All data is for
    one random realization of a chaotic dual-unitary gate arranged in the
    brickwall geometry \dif{($v=\vv=2$)}. (a) Eigenvalues $\lambda_j$ of $\Mr$ in the complex
    plane and their condition numbers $\kappa_j$ (colors). The
    number of nonzero eigenvalues is $6 \cdot 4^{(r - 1)/2} = 384$ and $\lambda
    = 0$ is $24192$ times degenerate. (b) Partial probability norms $w_s$
    (\ref{eq:ws}) diverging as $\sim \mu^s$ with $\mu \approx 1.38$. The
    vertical axis is scaled by $\kappa_1$ at the given $r$. The
    equality $|\lambda_1| \mu^2 = 1$ is satisfied up to machine precision. (c)
    The leading condition number $\kappa_1$ (\ref{eq:kappaj}) is staggered for
    odd/even support $r$, but still diverges with the same $\mu$ for each parity
    of $r$.}
    \label{fig:fig3}
\end{figure*}
We now turn to dual-unitary (DU) circuits, which are brickwall quantum circuits
composed of unitary gates that are unitary also in the spatial direction.
Namely, the propagator for one time step is defined as
\begin{align}
    U &= U_\mathrm{odd} U_\mathrm{even} , \label{eq:circ_prop}\\
    U_\mathrm{odd} &= V_{1, 2} V_{3, 4} \cdots V_{L - 1, L}, \nonumber \\
    U_\mathrm{even} &= V_{2, 3} V_{4, 5} \cdots V_{L - 2, L - 1} V_{L, 1}, \nonumber
\end{align}
where $V_{i, j}$ denotes the unitary $V$ acting on sites $i$ and $j$.
We can additionally define the spatial gate $\tilde V_{1, 2}$ by rearranging the indices
of $V_{1, 2}$
\begin{align}
	[\tilde V_{1, 2}]_{i,j;k,l} = \left[V_{1, 2}\right]_{j,l;i,k},
\end{align}
where $[\bullet]_{i,j;k,l}$ denotes the matrix elements between the tensor
product computational basis states $k, l$ and $i, j$ on the first two sites.
DU circuits are defined by the requirement that, in addition to $V$, $\tilde V$ is also unitary.

DU circuits were introduced in Ref.~\cite{bruno}, for a review see
Ref.~\cite{du_rev}. Because their dynamics is special due to the existence of both
an exact temporal and spatial causal cone, they allow for exact solutions of
certain quantities even in the quantum chaotic regime. In particular, their
two-point autocorrelation functions are nonzero only on the causal cone and their
behavior can be characterized with a simple transfer
matrix~\cite{bruno}.

Since DU circuits propagate local operators only on the causal cone, the
magnitudes of eigenvalues of their truncated propagator $\Mr$ have no $k$
dependence. Moreover, the $k= 0$ truncated propagator is exactly equal to the
local Heisenberg propagator aligned to the causal cone. In order to understand
its behavior, we shall study the graph of allowed operator transitions after the
application of one layer $\mathcal U_\mathrm{odd}$ followed by a translation $\mathcal T$ to the
left (to align to the causal cone~\cite{foot_du_align}) in the Heisenberg picture. The graph was
constructed in Ref.~\cite{holden_dye}, and a simplified version is reproduced in
Fig.~\ref{fig:du_graph}. Note that since the graph shows the allowed transitions
for half a period, $\M$ corresponds to two applications of the graph. For
generic quantum circuits, the graph would be completely connected, whereas for
DU circuits it is sparse and directed.

We observe that the graph has a stable fixed point, namely, all operators in
$\mathcal H_s$ (even-$j$, even-$r$) can propagate only back to $\mathcal
H_s$~\cite{foot_dugr}. Furthermore, the dynamics in $\mathcal H_s$
is simple, since it can only increase the support by $2$ -- it can
be interpreted as similar to the half-infinite shift operator $\sum_{n = 1}^\infty \ket{n +
2}\bra{n}$ \difa{(the shift-like dynamics in DU circuits was alluded to already
in Ref.~\cite{claeys_krylov} in the Krylov context)}. Since transitions to the same support $r$ are not allowed, potential
eigenvectors of $\Mr$ lying purely in $\mathcal H_s$ must have eigenvalues equal
to $0$ (analogous to the finite shift operator $\sum_{n = 1}^N \ket{n +
2}\bra{n}$, which has all eigenvalues equal to $0$).

Therefore, nonzero eigenvalues of $\Mr$ must correspond to eigenvectors with at
least some components in either $\mathcal H_1$ or $\mathcal H_2$ (the odd-$r$
subspaces). After the application of $\mathcal U$, a fraction of the weight in
say $\mathcal H_{1}$ transitions into $\mathcal H_s$, where it can then only
cascade to higher supports and not affect the eigenvalue. Such a cascade of
weights (i.e., partial norms) is numerically shown for a particular realization
of a dual-unitary circuit in Fig.~\ref{fig:fig3}(b) and will be discussed in more
detail in the following paragraph. The only other possible transition in
$\mathcal H_{1}$ is the transition back to $\mathcal H_{1}$ at the same support
$r$, which is thus the only process controlling the eigenvalue. This is one of the
crucial facts used (albeit in a different language) in Ref.~\cite{bruno} to
express the correlation functions of DU circuits as transfer matrices for each
support $r$ separately. Additionally, it implies that all eigenvalues of $\Mr$
converge immediately; once a nonzero eigenvalue $\lambda_j$ appears at some
$r_0$, it stays the same for all $r>r_0$. The spectrum of $\Mr$ for one generic
realization of a DU circuit, where the leading eigenvalue converges for $r_0 =
5$, is shown in Fig.~\ref{fig:fig3}. The number of nonzero eigenvalues is
precisely equal to the dimension of the odd-$r$ subspaces ($\mathcal H_1$ and
$\mathcal H_2$), while the degeneracy of the zero eigenvalue is equal to the
dimension of the even-$r$ subspaces ($\mathcal H_s$ and $\mathcal H_u$).

We now turn to deriving the structure of the nontrivial
(Gamow) eigenvectors exactly. Given such an eigenvector $\ket{\mathfrak R_i}$, let $r_0$ be the first support on which
it has nonzero components. As discussed in the previous paragraph, the components
there exactly determine the eigenvalue $\lambda_i$ and their structure depends on the particular
realization of the gate~\cite{foot_dutm}. For the example in Fig.~\ref{fig:fig3}, $r_0 = 5$, although
we can also observe other $r_0$, typically between $r_0 = 1$ and $r_0 = 7$.
For $r > r_0$, the weights then exactly cascade forward in a way that
satisfies Eq.~\eqref{eq:ineq}, which we now show~\cite{foot_duqc}.
Namely, any eigenvector by definition satisfies
\begin{align}
    \Mr \ket{\R_i} = \lambda_i \ket{\R_i},
\end{align}
where we take $r$ large. To determine
the partial norm, we now sum \dif{both sides of the equation} over all components with support $s$
\begin{align}
	 \sum_{\ba, \text{sup}\ba = s} |\bracket{\ba}{\Mr}{\R_i}|^2 &= \abs{\lambda_i}^2 w_s.
\end{align}
Provided that $s > r_0 + 4 {= r_0 + 2v}$ (recall that $\Mr$ corresponds to two applications of the graph in
Fig.~\ref{fig:du_graph}), all the components of the
eigenvector at $s - 4$ already lie in $\mathcal H_s$. Therefore, the matrix
elements on the LHS can only be non-zero for components of $\R_i$ with support
$s - 4$. Furthermore, taking into account that operators in $\mathcal H_s$ can transition only
back to $\mathcal H_s$ and unitarity,
\begin{align}
	w_{s - 4} = \abs{\lambda_i}^2 w_s.
\end{align}
In the scaling variables, $w_s \sim \mu^s$, assuming even $s$, and
using that $\vv = v = 2$ for brickwall DU circuits~\cite{du_vb}, we now obtain the equality
\begin{align}
	\abs{\lambda_i} \mu^\vv = 1.
\end{align}
The partial norm for the leading eigenvector of a particular realization is
shown in Fig.~\ref{fig:fig3}(b). We also observe that due to the cascade in
$\mathcal H_s$ increasing the support by $2$, all partial norms $w_s$ for odd
$s$ are $0$, except for $s = r_0$.

An analogous derivation can be made for the left eigenvectors $\ket{\L_i}$.
Since they are eigenvectors of time-reversed dynamics $\Mr^\dagger$,
their fixed point is $\mathcal H_u$ (odd-$j$ and even-$r$). Therefore,
for $s > r_0$, $\ket{\L_i}$ only have nonzero components in $\mathcal H_u$, meaning that the inner
product $\braket{\L_i}{\R_i}$ has nonzero contributions only at $s = r_0$. This,
in particular, means that only the partial norms determine the scaling of $\kappa_i$,
and it is trivially $\kappa_i \sim \mu^r$ with different prefactors for odd and
even $r$. We numerically confirm this in Fig.~\ref{fig:fig3}(c).

\dif{
\begin{figure*}[t!]
  \centerline{\includegraphics[width=\textwidth]{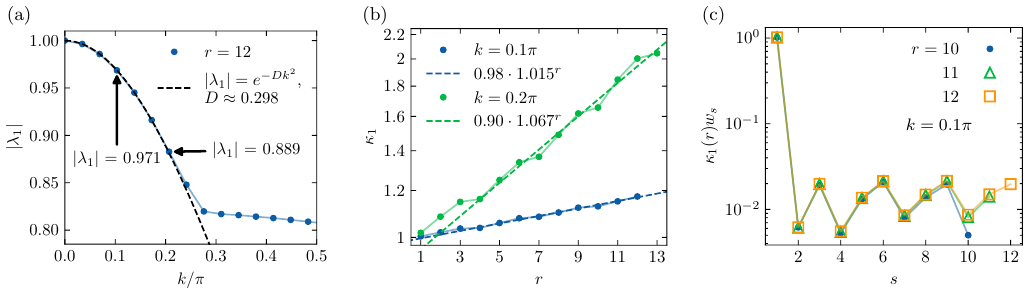}}
  \caption{\difa{\textbf{Fractal cascade in magnetization-conserving circuits.} All
  data is for a single realization of Eq.~\eqref{eq:3site} labeled
  ``Gate 1'' in Ref.~\cite{urban}. (a) Gaussian quasimomentum dependence of the
  leading eigenvalue $\lambda_1$ of the truncated propagator $\M_r$ and the
  fitted spin diffusion constant $D$. (b) Leading condition numbers $\kappa_1$
  diverge as $\sim \mu^r$ with visible 3-site staggering, data for two $k$
  denoted by arrows in (a). (c) Partial norms $w_s$ of the leading RP
  eigenvector at $k = 0.1\pi$. Rescaled by $\kappa_1$ to achieve data collapse
  for different $r$.}}
  \label{fig:u1}
\end{figure*}

\section{Conservation laws and higher dimensions}
\label{sec:beyond}

In Sec.~\ref{sec:main}, we explained the operator cascade in 1D chaotic systems
without conservation laws. In this section, we show how it extends to more general
settings, that is, to systems with conservation laws (Sec.~\ref{sec:u1}) and higher-dimensional
systems (Sec.~\ref{sec:2D}).

\subsection{Systems with conservation laws}
\label{sec:u1}

In order to study the operator cascade in the presence of conservation laws, we
consider a chaotic magnetization-conserving circuit. One of the simplest
possible examples that consists of the same gate everywhere is a 3-site
brickwall circuit (3-site interaction is needed to break
integrability~\cite{U1})
\begin{align}
    U = U_3 U_2 U_1, \qquad
    U_i = \prod_{j = 0}^{L/3} V_{i + 3j, i + 3j + 1, i + 3j + 2}, \label{eq:3site}
\end{align}
where $L$ is the number of sites, and both the three-site gate $V$ by itself and
the whole system $U$ conserve the magnetization $M = \sum_j \sigma_j^z$. This is
achieved simply by taking $V$ with an appropriate block structure in the computational
basis, see Fig.~3 in Ref.~\cite{urban}.

We studied RP resonances of such circuits in Ref.~\cite{urban}, where we showed
that the truncated propagator $\M_r$ is gapless at quasimomentum $k = 0$ with
the eigenvector being the magnetization $M$, while
its leading eigenvalue for small enough $k$ governs the decay of correlations of
the Fourier transform of local densities $\ave{\sigma^z(t, k) \sigma^z(0, k)}
\sim \e{-Dk^2 t}$~\cite{forster} and is equal to
\begin{equation}
	\abs{\lambda_1(k)} = \e{-Dk^2}, \label{eq:u1_lam_k}
\end{equation}
where $D$ is the spin diffusion constant, as demonstrated in
Fig.~\ref{fig:u1}(a). The spectrum of $\M_r$ for $k > 0$ is, therefore, once
again gapped as in chaotic systems without conservation laws and one expects the
corresponding eigenvectors to be singular by the same arguments as provided in
Sec.~\ref{sec:div_vec}. We indeed observe this, as demonstrated in
Fig.~\ref{fig:u1}(b, c). Note that the 3-site staggering is strongly present
(similar to generic quantum circuits in Appendix~\ref{app:circs}), with the data
being cleaner for the condition number $\kappa_1$, where we fit the divergence rate
$\kappa_1 \sim \mu^r$. While $w_s$ also increase
asymptotically with the rate depending on the 3-parity of $s$ ($w_{s + 3}/w_s >
1$ for large enough $s$), we do not fit the rate due to a small number of
accessible points (see Appendix~\ref{app:circs}, Fig.~\ref{fig:generic} for the same effect in circuits
without conservation laws). Importantly, the data for partial norms $w_s$ at
different $r$ collapses when rescaled with $\kappa_1(r)$, which implies that the
expansion coefficients between the leading RP eigenvector and local observables
$c_1$ \eqref{eq:cj} is finite in the TDL, as argued in Sec.~\ref{sec:div_vec}.

From $w_s$ of a time-evolved $\sigma^z$ operator with momentum $k > 0$ (in a
finite system of size $L = 14$ we can evolve only up to $t = 2$, when we already
hit the boundary), we estimate the butterfly velocity to be in the range $\vv
\approx 1.6\text{--}2.4$. Within this range, we observe the validity of
Eq.~\eqref{eq:ineq}, namely, $\abs{\lambda_1(k)} \mu^\vv \approx 0.99\text{--}1.04$ for
$k = 0.1\pi, 0.2\pi$ (Fig.~\ref{fig:u1}). Applying Eq.~\eqref{eq:u1_lam_k}, this
allows us to express the many-body fractal dimension in terms of the spin diffusion
constant
\begin{equation}
  \dt = Dk^2 \approx \dx \vv.
  \label{eq:Dm}
\end{equation}
While we expect the singular structure of eigenvectors to be valid in general
chaotic systems with conservation laws, because we expect that the gap to
subleading resonances closes in the TDL~\cite{urban}, it is not clear whether the
convergence of $\lambda_1$ is still exponential in $r$ and whether there are any
corrections to Eq.~(\ref{eq:Dm}).

In Appendix~\ref{app:pre}, we confirm the cascade also in the vicinity of the
Trotter limit of the kicked Ising model in which energy is conserved.
}
\dif{

\subsection{Higher-dimensional systems}
\label{sec:2D}

\begin{figure}[b!]
  \centerline{\includegraphics[width=3.3in]{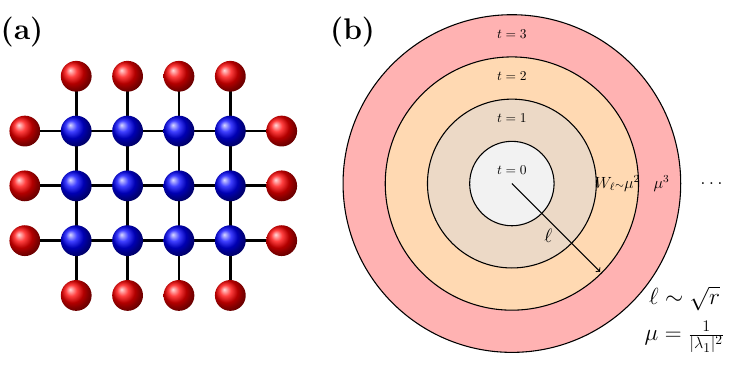}}
  \caption{\dif{{\bf Operator spreading in 2D.} (a) In 2D circuits with n.n.\ commuting gates the operator density at large time $t$ (blue) can spread in one time step to additional red sites. (b) Due to unitarity one has a scaling structure at large $t$: the total weight of operators in the next layer increases by $1/|\lambda_1|^2$.}}
  \label{fig:area2d}
\end{figure}
To derive how partial norms $w_s$ and condition numbers $\kappa_1$ diverge in
locally interacting systems in more than one dimension, it is best to remind
ourselves of the core reason for this behavior, which is easiest to see in the
derivation of the Eq.~(\ref{eq:ineq}). In $1=\nor{A(t+1)}^2/\nor{A(t)}^2$, one
has a $|\lambda_1|^2$ factor due to the decay of correlations, which must in
turn be compensated by the increased weight of newly populated operators. In
other words, at each time step the weight of freshly populated operator
densities has to increase by a factor $1/|\lambda_1|^2$, see
Fig.~\ref{fig:area2d}. This is transparently seen in Eq.~(\ref{eq:kappalam}) in
which the exponent $\frac{r}{2\vv}$ is nothing but the time needed for operator
densities to spread to size $r$. Generalization to $d$ spatial dimensions is now
clear: at long times (i.e., support number $r$), when the subleading
contributions are negligible, one has $|\lambda_1|^{2} W_{\ell+\vv}/W_{\ell}=1$,
where $W_{\ell}=\sum_{s \in {\rm ball}(\ell)} w_s$ is the weight of all
densities that are effectively reached in time $t$, i.e., densities in a ``ball'' that we label
by its radius $\ell$ at time $t$ ($\ell$ increases by $\sim \vv$ at every time
step), see Fig.~\ref{fig:area2d}(b). The truncation parameter $r$, i.e., the
maximal number of retained operators, will be equal to the volume of the
``ball'' (causal cone), which for a $d$-ball of radius $\ell=\vv t$ is
$\pi^{d/2} \ell^d/\Gamma(d/2+1)$. For a specific lattice, the prefactor will be
different from the one for a ball and we simply write $r=(ct)^d$. This results in an
exponential growth of $W_\ell$ with $\ell$, or, equivalently, in an exponential
growth with $r^{1/d}$, leading to
\begin{equation}
  \kappa_1 \sim \left( \frac{1}{|\lambda_1|^2} \right)^{\frac{r^{1/d}}{c}}, \quad w_s \sim \left(\frac{1}{|\lambda_1|^2} \right)^{\frac{s^{1/d}}{c}}.
  \label{eq:2dkappa}
\end{equation}
Here we have already assumed that we are in the asymptotic regime of
sufficiently large $r$ such that the equality (\ref{eq:ineq}) holds and one can
express $\mu$ by $\lambda_1$. Compared to 1D, there is an important difference;
the divergence is exponential in the linear size $\ell=r^{1/d}$ of the
truncation and not in the maximal total number of non-identity operators $r \sim
\ell^d$ kept in the truncation. For instance, if in 2D one truncates to all
extensive operators whose density is supported on a $3\times 3$ lattice patch, that is,
one has $r=9$, the linear size is $\ell=3$. Note that one still has a
multiplicative scenario that led us to define the many-body fractal
dimension $\vv \dx=\vv \ln{\mu}=-\ln|\lambda_1|$ in 1D, which tells us the asymptotic
scaling between consecutive layers formed in 1D by operators with density
supported on $s$ sites. In $d$-dimensions, one has onion-like layers formed by
operators with different ``diameters'' $\ell=r^{1/d}$ reached at different
times, see Fig.~\ref{fig:area2d}(b). The scaling factor, i.e., the fractal
dimension, is again given by $\lambda_1$.

Let us numerically check the above prediction in 2D, specifically that the growth is now exponential in $\ell = \sqrt{r}$, if $r$ is the total number of operators kept in the truncation (i.e., $r$ is proportional to the area). We take a 2D kicked Ising model on a 2D square lattice, with a one-step propagator being
\begin{eqnarray}
  U_\tau&=& U'_{\rm zz}  U'_{\rm z} U'_{\rm x},\qquad U'_{\rm zz}= \e{-\ii \tau \sum_{\rm n.n.} \sz_{i,j} \sz_{i',j'}},\\
  U'_{\rm x}&=&\prod_{i,j} \e{-\ii \tau \hx \sx_{i,j}},\quad U'_{\rm z}= \prod_{i,j} \e{-\ii \tau \hz \sz_{i,j}}, \nonumber
  \label{eq:2D}
\end{eqnarray}
where the sum in $U'_{\rm zz}$ is over nearest neighbors of a 2D square
lattice, and we use $\hx=\hz=1$. The 2D Ising model has been the subject of
several recent quantum computer experiments~\cite{IBMutility23,feig25}. The
truncated propagator is constructed in a way similar to that in 1D systems, with
details provided in Ref.~\cite{prx}. Because we can afford to simulate only
modest linear truncation sizes $\ell$, specifically we check 4 different $r$
ranging from $r=2\times 2=4$ truncation with $\ell=2$ to $r=3\times 3$ with
$\ell=3$ (see small diagrams in Fig.~\ref{fig:2Dkappa1}), we in addition checked
several different values of gate duration $\tau$ to nevertheless be able to
better judge if the scaling $\sqrt{r}$ agrees with data. Results are shown in
Fig.~\ref{fig:2Dkappa1}. We can see that the data fits the theory (\ref{eq:2dkappa})
rather well, in which we use empirical $\lambda_1=0.176,0.30,0.47,0.78$ for
$\tau=0.70,0.65,0.60,0.50$, respectively, and the only fitting parameter is
$2/c=0.5$ in the exponent (the overall prefactor is always $1$). In
Appendix~\ref{app:2D}, we also demonstrate that exponential divergence of
$\kappa_1$ with $\sqrt{r}$ fits the data much better than divergence with $r$
(as in 1D).
\begin{figure}[t!]
  \centerline{\includegraphics[width=3.0in]{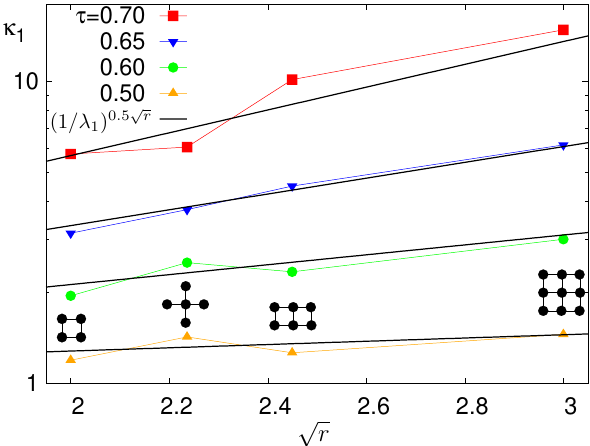}}
  \caption{\dif{{\bf Fractal cascade in 2D.} 2D kicked Ising model (\ref{eq:2D}) and the exponential
  divergence of $\kappa_1$ with $\sqrt{r}$ for the leading RP resonance. Due to
  the limited range of truncation sizes $r$ (diagrams with black circles) we also show data for different gate durations $\tau$.}}
  \label{fig:2Dkappa1}
\end{figure}

}

\section{Additional spectral properties of the truncated propagator}
\label{sec:math}

In this section, we discuss some additional spectral properties of the truncated
propagator. In Sec.~\ref{sec:conv}, we first argue that the convergence of RP
resonances is exponential in support $r$, which makes the method numerically
attractive, and connect the convergence to partial binorms. In
Sec.~\ref{sec:kappa}, we study the properties of the noisy non-physical part of
the spectrum, in particular condition numbers, and discuss whether the pseudospectrum
of $\Mr$ is relevant.

\subsection{Exponential convergence of Ruelle-Pollicott resonances}
\label{sec:conv}

\begin{figure}[t!]
  \centerline{\includegraphics[width=3.0in]{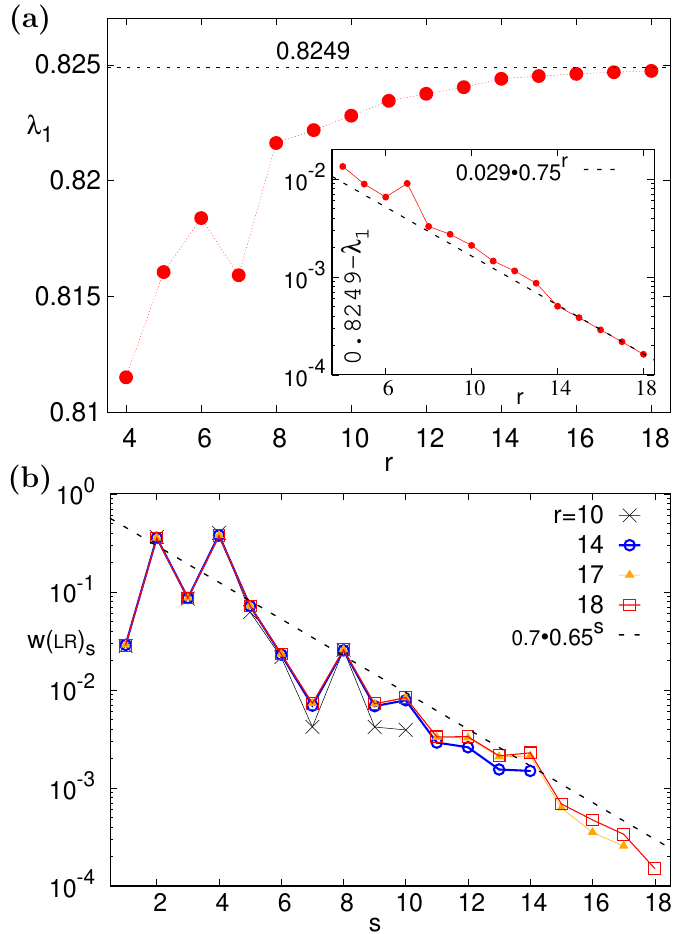}}
  \caption{{\bf Convergence with $r$.} (a) The largest RP resonance converges
  exponentially with the truncation $r$. (b) Partial binorms $w_s^{(\L\R)}$
  (\ref{eq:bin}) of the leading left and right eigenvectors
  ($\braket{\L_1}{\R_1}=1$) exhibit roughly exponential dependence on $s$.
    Both panels show data for the 1D kicked Ising model at $\tau=0.75, k=0$, and standard $\hx=0.9,
  \hz=0.8$.}
  \label{fig2}
\end{figure}

While RP resonances have been observed to converge for numerically accessible
$r$~\cite{RP24, urban,prx}, it was not clear whether the convergence is exponential. Due
to improved numerical data, we now claim that convergence is generically
exponential, as seen in Fig.~\ref{fig2}(a) for the kicked Ising model. Exponential
decay with a similar rate is observed also for partial binorms defined as
\begin{align}
  w^{(\L \R)}_s = \sum_{\bb, \text{sup}(\bb)=s} \tilde b_{\bb}^* b_{\bb}, \label{eq:bin}
\end{align}
where $b_{\bb}=\braket{\bb}{\R_1}$ and $\tilde b_{\bb}^*=\braket{\L_1}{\bb}$ are
expansion coefficients in the local operator basis, analogous to the definition
of partial norms in Eq.~\eqref{eq:ws}. Their exponential decay is shown
in Fig.~\ref{fig2}(b) and was previously
already observed in Ref.~\cite{Prosen}. In this section, we shall argue that the
decay of binorms is a physical consequence of the fact that the probability of
backflow processes, defined as processes where operators with long support $s$ shrink back to
some small support, is exponentially suppressed in $s$. Exponential convergence
of RP resonances is then related to the exponential decay of partial binorms.

Let us first discuss the behavior of partial binorms. As argued already in
Sec.~\ref{sec:fractal}, a generic local operator will evolve towards
$\ket{\R_1}$ for positive time and towards $\ket{\L_1}$ for negative time. We choose a
Pauli string $\ket{\ba}$ with the smallest possible support that has overlap
with both $\ket{\R_1}$ and $\ket{\L_1}$, and normalize $\ket{\R_1}, \ket{\L_1}$ and $\ket{\ba}$ in such a way that
$\braket{\L_1}{\ba} = \braket{\R_1}{\ba} = \braket{\L_1}{\R_1} = 1$. We, therefore, have $\ket{\R_1}
\approx \frac{1}{\lambda_1^t} \Mr^t \ket{\ba}$ and $\bra{\L_1} \approx
\frac{1}{\lambda_1^{t}} \bra{\ba}\Mr^{t}$ for large enough $t$. The
partial binorm can now be expressed as
\begin{align}
	w_s^{(\L \R)} &= \sum_{\bb, \mathrm{sup}(\bb) = s} \braket{\L_1}{\bb} \braket{\bb}{\R_1} \nonumber \\
	&\approx \frac{1}{\lambda_1^{2t}}\sum_{\bb, \mathrm{sup}(\bb) = s} \bracket{\ba}{\Mr^t}{\bb} \bracket{\bb}{\Mr^t}{\ba}, \label{eq:backflow}
\end{align}
for $r \gg s \gg 1$. The sum is taken over a basis of Pauli strings $\bb$ with
support $s$. The sum in Eq.~\eqref{eq:backflow} is the probability
amplitude for $\ba$ to exhibit a kind of ``backflow echo'', i.e., first expanding to support $s$ in time $t$ and
then shrinking back to the initial support in the same time $t$.

Backflow processes in quantum circuits were studied in Ref.~\cite{tibor22b}. We
now briefly repeat their argument that implies the exponential suppression of
$w_s^{(\L \R)}$ in $s$. We first square Eq.~\eqref{eq:backflow}, and then assume that the phases in the sum are essentially random (reasonable for chaotic systems and large $t$)
\begin{align}
	\hspace{-9pt}\abs{w^{(\L \R)}_s}^2 \hspace{-3pt} &= \hspace{-1pt} \frac{1}{\abs{\lambda_1}^{4t}}\sum_{\bb, \bb'} \bracket{\ba}{\Mr^t}{\bb} \bracket{\bb}{\Mr^t}{\ba} \bracket{\ba}{\Mr^t}{\bb'}^* \bracket{\bb'}{\Mr^t}{\ba}^* \nonumber \\
	&\approx \frac{1}{\abs{\lambda_1}^{4t}}\sum_{\bb, \mathrm{sup}(\bb) = s} |\bracket{\ba}{\Mr^t}{\bb}|^2 |\bracket{\bb}{\Mr^t}{\ba}|^2,
        \label{eqx}
\end{align}
where the sums in the first line run over $\mathrm{sup}(\bb) = \mathrm{sup}(\bb')=s$.
Unitarity implies $\sum_{\bb} |\bracket{\bb}{\Mr^t}{\ba}|^2 \approx 1$ and
thus $\sum_{\bb, \mathrm{sup}(\bb) = s} |\bracket{\bb}{\Mr^t}{\ba}|^2 \leq 1$.
We now replace the 1st factor in the 2nd line in Eq.~\eqref{eqx} with a
factor obtained for a specific $\bb_{\rm max}$ (of support $s$) for which
$|\bracket{\ba}{\Mr^t}{\bb}|^2$ is maximal and use unitarity in the remaining
sum, resulting in
\begin{align}
  	\abs{w^{(\L \R)}_s} \lesssim \frac{1}{|\lambda_1|^{2t}}|\bracket{\ba}{\Mr^t}{\bb_{\rm max}}|. \label{eq:pbn_bound}
\end{align}
The time-dependent prefactor simply
fixes the normalization and cancels out in an exact calculation. In
random unitary circuits, the RHS of Eq.~\eqref{eq:pbn_bound} can be evaluated
exactly~\cite{curt18}, with a time-independent part, relevant for us, being
simply $2^{-s}=1/\sqrt{N_s}$ due to a counting argument. This would mean that the
partial binorm decays as
$\eta^s = 0.5^s$ or faster. Both the kicked Ising model and
brickwall circuits in Appendix~\ref{app:circs} have more
structure than random circuits, namely translational invariance and homogeneity
in time, which cause the decay of binorms to be slower. We observe $\eta \approx 0.65\text{--}0.75$ in all cases.
\difa{\mz{A similar $\eta \approx 0.5\text{--}0.65$ is also observed} in the magnetization-conserving
model \mz{studied} in Sec.~\ref{sec:u1} (data not shown). See also Ref.~\cite{tibor22b} for an in-depth
discussion of backflow in magnetization-conserving circuits.}

As discussed in Sec.~\ref{sec:du}, the decay of binorms is immediate in
dual-unitary circuits, which is a consequence of merely forward-flowing
shift-like dynamics in the cascading subspace $\mathcal H_s$, i.e., the absence
of backflows there. The above discussion for general circuits can now be
interpreted in a similar way asymptotically. That is, one can imagine the
dynamics of local operators in generic many-body systems as converging to
shift-like dynamics at large support.

Finally, we discuss the numerical observation that the convergence rate of RP
resonances is similar to the decay rate of partial binorms. The connection can
be illuminated by a heuristic argument. For some large $r$, the contribution of
supports larger than some $s$ (keeping $s \ll r$) to the eigenvalue can be
approximated as $\Delta \lambda(s) \approx \bracket{\L_1}{\mathcal P_{>s} \Mr
\mathcal P_{>s}}{\R_1}$, where $\mathcal P_{>s}$ is the projector to the space
of all operators with support larger than $s$. Due to $\mathcal U_r$ coupling
only a finite number of consecutive supports $2v$, $\mathcal P_{>s}\ket{\R_1}$
is still almost an eigenvector of $\mathcal U_r$ with an eigenvalue
$\abs{\lambda_1} < 1$. This finally allows us to make an estimate $\abs{\Delta
\lambda_1(s)} \lesssim \abs{\bracket{\L_1}{\mathcal P_{>s}}{\R_1}} =
\abs{\sum_{s' = s + 1}^r w_{s'}^{(\L \R)}} \sim \eta^s$. In other words,
eigenvalues converge in the same exponential way as the partial binorms.
Providing a rigorous version of the presented perturbation-theory-like argument
(see also Appendix~\ref{app:pert}) is an interesting open question.

\begin{figure}[t!]
  \centerline{\includegraphics[width=2.7in]{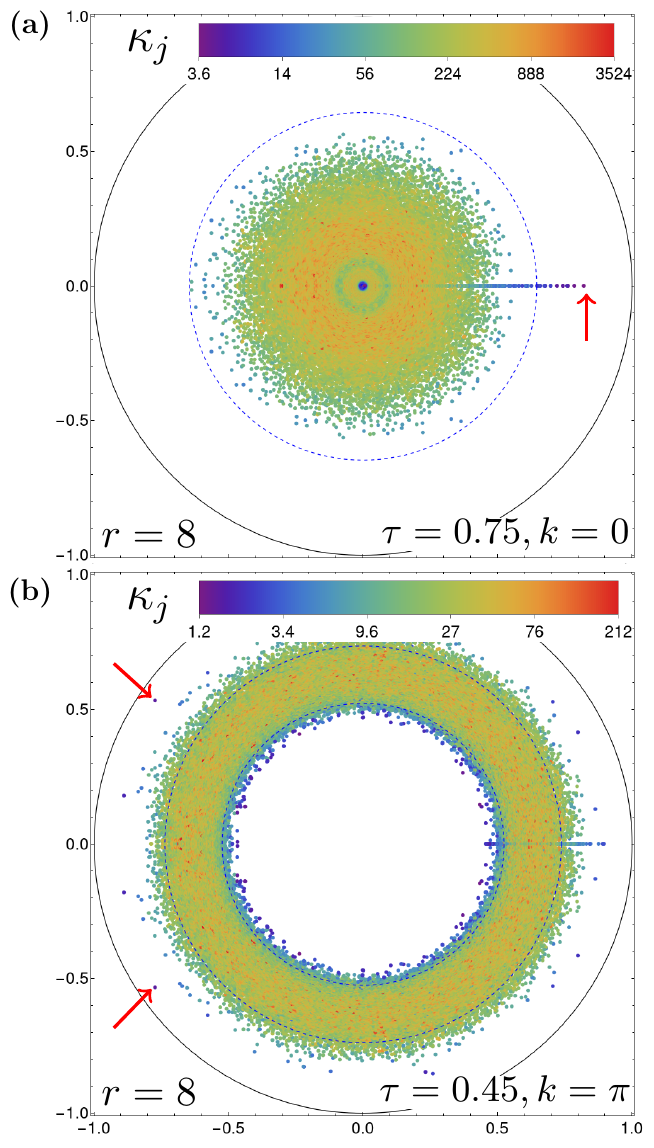}}
  \caption{{\bf Condition numbers}. Eigenvalues $\lambda_j$ of $\Mr$ in the complex
  plane and their condition numbers $\kappa_j$ (colors). $\kappa_j$ of
  eigenvalues in the bulk, i.e., within the two rings given by
  Eq.~(\ref{eq:sr_r}) (dotted circles), are much larger than $\kappa_1$ (red
  arrows). 1D kicked Ising model for $\hx=0.9$, $\hz=0.8$, and $\tau=0.75, k=0$ in
  (a), and $\tau=0.45, k=\pi$ in (b). Convergence of $\lambda_1$ with truncation
  $r$ is shown in Fig.~\ref{fig2}(a) for data in (a), and in Fig.~\ref{fig:fig9}(a)
  for data in (b).}
  \label{fig:fig4}
\end{figure}

\subsection{Noisy bulk of the spectrum}
\label{sec:kappa}

The spectrum of the truncated propagator generically consists of two parts, a
noisy bulk and a few isolated eigenvalues outside of it, which we interpreted as
RP resonances (Fig.~\ref{fig:fig4}). In this section, we shall discuss the differences
between properties of the two sets of eigenvalues.

One obvious difference is the physical separation of $\lambda_j$ in the complex
plane. In Ref.~\cite{RP24}, it was observed that the location of the noisy bulk
can be modeled with a random matrix ensemble $U \Sigma V$, where $U, V$ are Haar
random unitary matrices and $\Sigma$ is a diagonal matrix of singular values of
the considered $\Mr$. The single ring theorem~\cite{Zee,ring} then states that
the eigenvalues are concentrated in an annulus with radii determined by the
averages of singular values
\begin{align}
	r_\text{in} = 1 / \sqrt{\ave{\Sigma_j^{-2}}}, \qquad r_\text{out} = \sqrt{\ave{\Sigma_j^2}}. \label{eq:sr_r}
\end{align}
Using the conjectured exact values of the singular values of $\Mr$ for the kicked
Ising model~\cite{RP24}, we can see in Fig.~\ref{fig:fig4} that the ring radii
are indeed roughly given by Eq.~(\ref{eq:sr_r}). Note that this noisy ring-like
spectrum is very different from the spectrum obtained by simply projecting a
unitary to a random subspace (taking a block of a unitary matrix)~\cite{karol}
-- physically motivated truncation by the operator support is crucial for the
properties of $\Mr$. The singular values turn out to be the same for any
truncation length $r$, only their multiplicities increase exponentially in $r$. As discussed in Appendix~\ref{app:circs}, we see a similar separation
in the spectrum of $\Mr$ for generic brickwall circuits, but with singular
values in that case being different for even and odd $r$.
In dual-unitary circuits, the noisy bulk is absent due to each eigenvalue converging immediately at a certain $r_0$, as argued in Sec.~\ref{sec:du}.

The separation between RP resonances and the noisy bulk can also be seen by
studying the condition numbers $\kappa_j$. As seen in Fig.~\ref{fig:fig4}, the
condition numbers in the bulk are much larger than those of eigenvalues outside
of the bulk. Specifically, we observe that the median $\kappa_j$ (i.e., in the
bulk) scales approximately as $\sim \sqrt{N} \sim 2^r$, where $N$ is the
dimension of the matrix $\Mr$, whereas $\kappa_1$ of the leading RP resonance
increases much more slowly (empty red circles vs. full red circles in
Fig.~\ref{fig1}(b)). The $\sim \sqrt{N}$ scaling is consistent with the exact
scaling of condition numbers in various random matrix theory (RMT)
ensembles~\cite{edelman,chalker,yan,belinschi}. We also note that there has been
some interest in the RMT community in the properties of RMT matrices subject to
a low-rank perturbation. A general scenario is that, for a sufficiently strong
perturbation, there will be outliers in the spectrum (akin to our RP resonances)
that have properties distinct from the
bulk~\cite{baik,vinayak,tao,rochet,potters}. An open question is whether those
results can be of use in understanding properties of $\Mr$, and if one could
perhaps devise a method that would filter out the ``noisy'' part of $\Mr$,
leaving us only with the bare RP resonances, making their numerical study
simpler.

Very large condition numbers are, in principle, an indication that the spectrum is
unstable. In such situations, it is better to look at a stable generalization
called the pseudospectrum~\cite{trefethen} (in a nutshell, a union of spectra of
all possible slight perturbations of a matrix). This is not just a remark on
stability of numerically computing $\lambda_j$, but it can in fact happen that
the pseudospectrum will give the correct decay rate while the spectrum will not.
Recently a number of such situations has been identified in
physics~\cite{okuma,viola,random,boom,tibor,ash,naves}, but in all cases
the condition numbers grew exponentially with matrix size $N$. In our case, $\Mr$
is only very weakly non-normal (especially $\lambda_1$) and we do not have to
worry about differences between the spectrum and the pseudospectrum of $\Mr$. \dif{As a side note, we recently nevertheless identified a specific case of abstract CMV matrices where the eigenvalues of the truncated propagator do not give the correct RP resonances~\cite{CMV}.}

One might also wonder if the spectrum of the truncated propagator $\Mr$ is in
fact equal to the pseudospectrum of the unitary ${\cal U}$, the argument being
that $\Mr$ for $r \to \infty$ is a ``small'' perturbation (truncation) of ${\cal
U}$~\cite{footpsU}. This, however, is not true. The truncated propagator $\Mr$
is a ``small'' perturbation of ${\cal U}$ only for {\em local operators}, i.e.,
in a physical sense. Mathematically, the difference $\Mr-{\cal U}$ is always
large no matter what finite $r$ we use (we cut off infinitely many operators
with a support larger than $r$).

\section{Discussion}

\dif{
By studying spectral properties of the operator propagator $\Mr$ truncated to
densities supported on at most $r$ sites, we reveal an intricate fractal
structure that explains how relaxation, and thereby chaos, arise in quantum
many-body systems without any classical limit, for instance, in quantum circuits
of high interest in quantum simulations. In particular, all the ingredients of
the asymptotic expression $\Mr^t \approx \lambda^t_1
\ket{\R_1}\bra{\L_1}/\braket{\L_1}{\R_1}$ have a clear meaning: $\lambda_1$
gives the decay of correlation functions, the diverging self-similar structure
of eigenvectors $\ket{\R_1}$ and $\ket{\L_1}$ explains how quantum many-body
chaos comes about, while the exponential decay of partial mutual overlaps, i.e.,
partial binorms, tells us that the numerical convergence with $r$ is
exponential.}

The self-similar many-body hierarchical structure of eigenvectors bears some similarities to stable and unstable manifolds of classical chaotic
systems, which are responsible for chaos and the difficulty in simulating such
systems. Due to its conceptual similarity to turbulence in classical fluids,
where energy flows from large eddies at short times to increasingly smaller
eddies at later times, we call the scenario the many-body operator Kolmogorov
cascade. At long times, relaxation in quantum many-body systems appears because
local operators flow to increasingly non-local many-body ones.
The rate
of increase of non-locality -- the many-body fractal dimension $\dx$ -- is encoded in the
exponential scaling of the partial norms of the left and the right
eigenvector (or, equivalently, of the condition number) with the operator's
\dif{linear support size $\ell \sim r^{1/d}$ in $d$ dimensions. The cascade is
generic in chaotic systems; it can be found in more than one-dimensional
systems, as well as in systems with symmetries. We have explicitly checked quantum
circuits with magnetization as well as energy conservation.}

In fluid turbulence, the conservation of energy and momentum is crucial and
leads to a certain universality in the value of scaling exponents. In quantum
chaotic systems without any local conservation laws, we only have conservation
of probability, i.e., unitarity, which nevertheless puts a constraint on the
fractal dimension. While its value is system-specific, it is related to the
decay of correlation functions $C(t) \sim \e{-\dt t}$ by an asymptotic equality
\difa{$\dx \vv \approx \dt$, with $\vv$ being the effective operator causal
velocity (the so-called butterfly velocity).} \dif{In systems with a conserved
quantity, we on the other hand find that the fractal dimension $\dx$ is related
to the diffusion constant $D$.}

An important insight is provided by exact results for dual-unitary circuits.
Each Ruelle-Pollicott (RP) eigenvalue has a characteristic operator density
support $r_0$ beyond which the structure of eigenvectors is exactly that of a
shift -- an exact multiplicative flow only to higher supports. Generic circuits,
on the other hand, approach such a pure-shift picture only asymptotically for
isolated RP resonances at a sufficiently large truncation $r$. \dif{This
explains when and why the truncated propagator method works. Provided the
truncation $r$ is large enough so that the leading RP resonance converges (in
systems without conservation laws the convergence is exponential), it will
accurately predict the behavior of correlation functions of local
observables, but, due to a diverging nature of eigenvectors, not of the whole evolution of the operator $\ket{A(t)}$. Diverging eigenvectors can also be rigorously treated within the rigged Hilbert space formalism; for
details see the exactly solvable cases in recent Ref.~\cite{CMV}. We, therefore, offer a broader picture of various Pauli string
truncation methods introduced in the last couple of years, and why it pays off
to work directly in the thermodynamic limit and explicitly resolve any of the
symmetries, like we do.}

The shift-like dynamics in the many-body operator space is suggestive of
the symbolic dynamics description of the so-called B-systems (Bernoulli systems)
that are at the top of the classical ergodic hierarchy of dynamical systems
(B-systems $\subset$ Kolmogorov systems $\subset$ mixing $\subset$ ergodic). How
far this analogy can be pushed, i.e., if the structures we revealed can indeed
serve as a quantum analog of the Markovian symbolic dynamics description in
classical systems, remains to be explored~\cite{footB}. In relation to classical
chaos, we can also highlight one conceptual difference. \dif{While in classical
few-degree-of-freedom system chaos appears due to an infinite-dimensional space of
functions describing evolution of densities with ever finer details, in our case
the infinity comes due to the thermodynamic limit. Relaxation and chaos, seen as
an escape to infinitely many-body operators rather than a stretch-and-fold
mechanism, arise from infinitely many degrees of freedom. Perhaps a scenario
like here -- complexity due to a spatial ``many-bodyness'' -- can be relevant
also in much less studied classical many-body systems.}

One of the core ideas of our approach is to focus on observables, specifically,
operators with local density, which are relevant experimentally. For instance,
the scaling of partial norms is encoded in the size of asymptotic correlations
(the prefactor) for observables with increasing support. For good choices of
circuits where the convergence to the asymptotic regime happens already at small
support $r_0$ (e.g., appropriate dual-unitary circuits), it might be within
reach of present-day quantum computers. \dif{Another route would be to design
protocols that would directly target the norm of the projector $\P_1$, i.e., the
condition number $\kappa_1$, which often converges with $r$ to its asymptotics
faster.} \difa{While we have focused on two-point correlation functions because
they are the easiest to measure and analyze theoretically, a study of quantities
involving more than one evolved operator, for instance, $O(t) =
\ave{A(t)BA(t)B}$, is possible. Because of the two (or more) propagators $\Mr$
involved, overlaps between eigenvectors become important and one might get
effects beyond the expected leading behavior $O(t) \sim \lambda_1^{2t}$.
Therefore, studying algebraic properties of eigenvectors of $\Mr$ is an intriguing open
question, especially since such higher-point correlations appear in
out-of-time-order correlators (OTOCs) and are of high experimental
interest~\cite{google}.}

\begin{center}
	\textbf{Acknowledgments}
\end{center}
UD would like to thank Rustem Sharipov, Pavel Orlov, Pavel Kos and Neil Dowling for insightful
discussions. M\v Z would like to thank Lucas S\' a for discussions. The authors
acknowledge Grants No.~J1-4385, No.~J1-70049, No.~N1-0504, and No.~P1-0402 from Slovenian
Research Agency (ARIS).

\clearpage

\appendix

\section{Additional data for the kicked Ising model}
\label{app:KI}

\begin{figure}[ht!]
  \centerline{\includegraphics[width=2.9in]{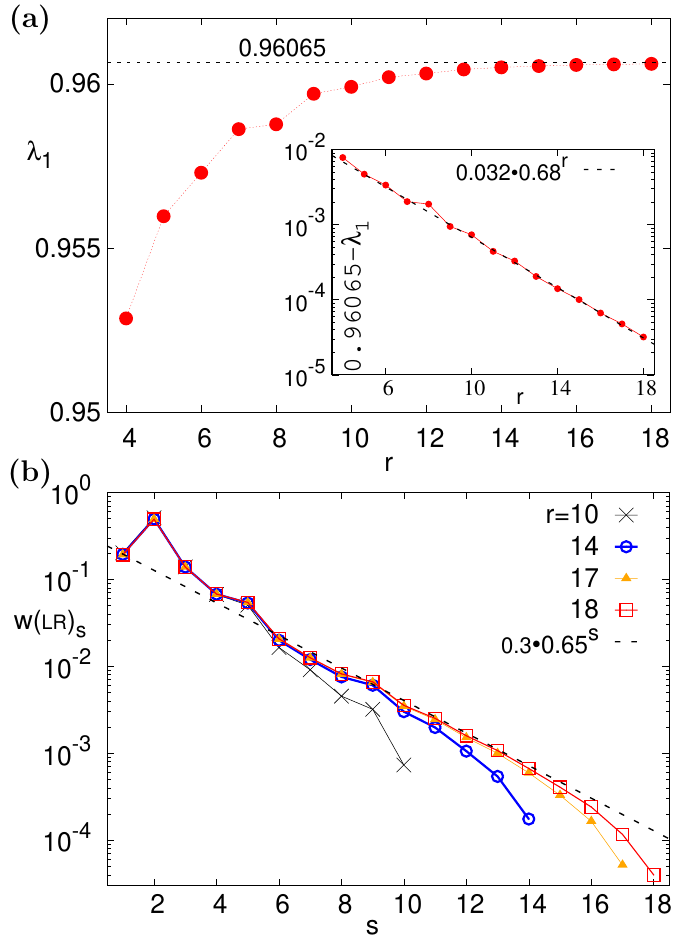}}
  \caption{{\bf Convergence with $r$.} (a) The largest RP resonance converges exponentially with the truncation $r$. (b) Partial binorms also exhibit an exponential dependence on $s$ with approximately the same rate as found in (a). All data is for the 1D kicked Ising model and $\tau=0.65, k=0$ ($\hx=0.9, \hz=0.8$).}
  \label{fig:fig2b}
\end{figure}
In Fig.~\ref{fig:fig2b}, we show that, similarly as for $\tau=0.75$ in
Fig.~\ref{fig2}, the convergence with the truncation size $r$ is exponential
also at $\tau=0.65$, with a similar rate (the factor $0.68^r$) as for
other parameters despite the gap $1-\lambda_1$ being rather different. We observe
similar results also for $\tau=0.65$ and $k=\pi$, where the two largest
$\lambda_{1,2} \approx -0.81 \pm \ii\, 0.21$ form a complex pair (data not shown).

\subsection{Multifractality}
\label{app:KI_mult}

\begin{figure}[t!]
  \centerline{\includegraphics[width=2.9in]{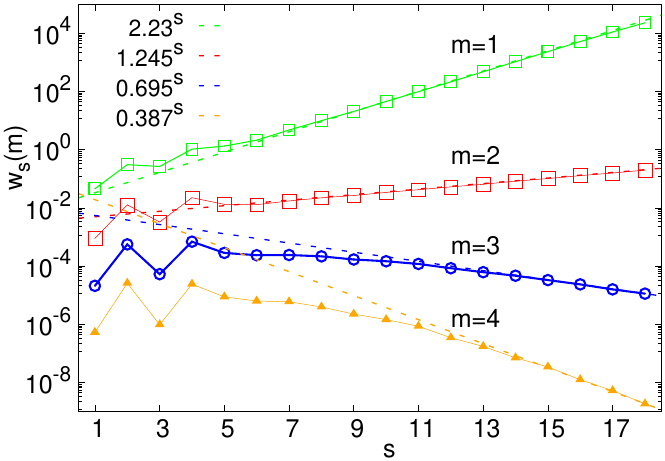}}
  \caption{{\bf Multifractality.} Absence of multifractality in different moments $m$ of $\R_1$ in the kicked Ising model at $\tau=0.75, k=0$, and $\hx=0.9, \hz=0.8$. Fitted exponents (dashed lines) agree with a prediction for ordinary fractals in Eq.~(\ref{eq:mum}) using $\mu=1.245$ (Fig.~\ref{fig1}).}
  \label{fig:fig2d}
\end{figure}

We have checked for the presence of multifractality in the 1D kicked Ising model by calculating partial norms of $\ket{\R_1}$ for a few exponents $m$,
\begin{equation}
  w_s(m) = \sum_{\ba,{\rm sup}(\ba)=s} |b_{\ba}|^m.
  \label{wsm}
\end{equation}
Due to fractality, one asymptotically expects $w_s(m)\sim \mu^s(m)$, and correspondingly an $m$-dependent dimension $\dx(m) := \ln{\mu(m)}$. In the main text we have focused on $\dx=\dx(2)$. Here, we also verify a few other powers $m=1,3$ and $4$. Multifractality would be indicated by a nonlinear dependence of $\dx(m)$ on $m$, while a linear dependence of $\dx(m)$, equivalently
\begin{equation}
  \mu(m)=4(\mu/4)^{m/2},
  \label{eq:mum}
\end{equation}
indicates an ordinary fractal. Numerical data for the 1D kicked Ising model is shown in Fig.~\ref{fig:fig2d} and suggests that there is no multifractality as all exponents agree with Eq.~(\ref{eq:mum}). \dif{An interesting open question is whether one can get any multifractality at some critical point, where the dynamics would exhibit a sudden change.}

\subsection{Non-scaling form of $w_s$}

\begin{figure*}[t!]
  \centerline{\includegraphics[width=0.97\textwidth]{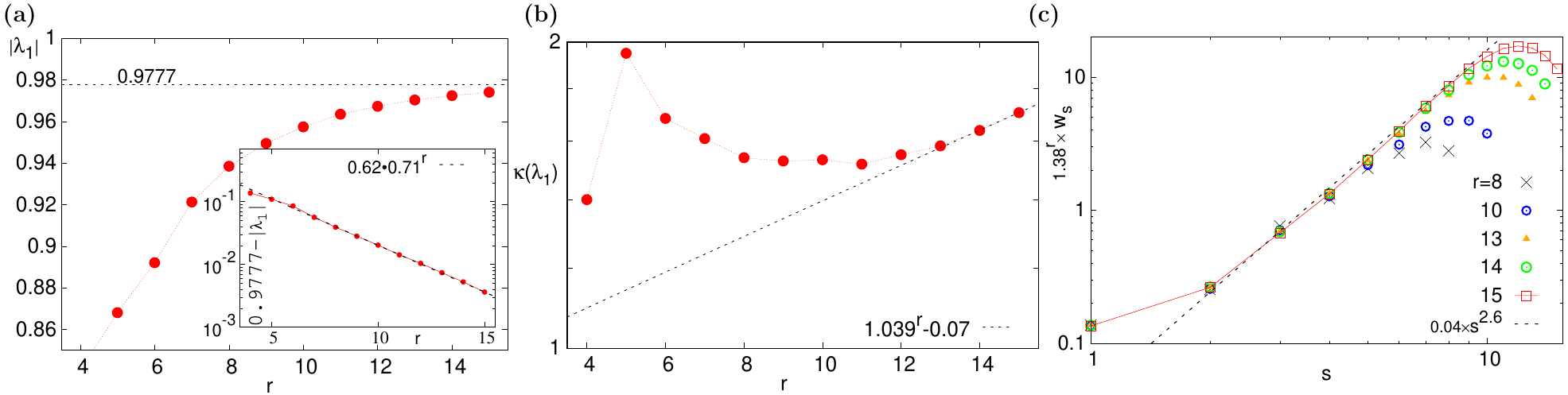}}
  \caption{\dif{{\bf Different scaling of $w_s$ and $\kappa_1$}}. 1D kicked Ising model at $\tau=0.45, k=\pi$ ($\hx=0.9, \hz=0.8$) where there are two complex largest eigenvalues, see Fig.~\ref{fig:fig4}(b). (a) Convergence of the two largest eigenvalues is exponential in $r$. (b) Condition number of the leading eigenvalue. (c) Partial norms $w_s$ of $\ket{\R_1}$ this time do not scale in the same way as $\kappa_1$.}
  \label{fig:fig9}
\end{figure*}
The situation is interesting at $\tau=0.45$ and $k=\pi$, that is, at large
quasimomentum where dynamics is not related to prethermalization and an almost
conserved energy found for small $k$ (see Appendix~\ref{app:pre}). The largest
eigenvalue $\lambda_1$ forms a complex pair (Fig.~\ref{fig:fig4}(b)) and
converges nicely exponentially in $r$ (Fig.~\ref{fig:fig9}(a)). The leading
condition number also suggests that it will grow exponentially as $\kappa_1 \sim
\mu^r$, with $\mu\approx 1.039$, Fig.~\ref{fig:fig9}(b) (though only the last
three $r$ agree with that), and the equality is satisfied, $|\lambda_1| \mu^\vv \approx 0.997$
($\vv \approx 0.5$). However, what is different is the scaling of partial norms
$w_s$, shown in Fig.~\ref{fig:fig9}(c). Two important observations are in order:
(i) to achieve an approximate collapse of data for different $r$ and $s$, one has
to use a large scaling factor $\approx 1.38$ on the vertical axis, and (ii)
growth does not seem to be exponential at large $s$ (the whole ``collapse'' is
much worse and it is hard to judge whether the growth is algebraic or
exponential in $s$). We observed before that often $\kappa_1$ converges in
a nicer way than $w_s$; however, in previous cases $w_s$ did eventually grow in
the same $\mu^s$ way, which does not seem to be the case here \dif{(or perhaps one needs still larger $r$ than $r=15$ available in this case)}. The fact that
the scaling in $w_s$ has a larger factor $1.38$ than in $\kappa_1$ means that
the coefficient $c_1$ goes to zero in the limit $r \to \infty$ (we have
explicitly checked that $c_1$ for $\ba=z$ decreases with $r$). If one
numerically calculates an appropriate correlation function from the sector
$k=\pi$, for instance of the staggered magnetization $S=\sum_j (-1)^j \sz_j$, it
is not clear if the asymptotic decay is algebraic or exponential -- for
a finite system size $L=32$, numerical data for $C(t)$ can be found in
Ref.~\cite{RP24}. A matrix product operator calculation in the thermodynamic
limit but with a matrix bond size $\chi=2048$ was also not conclusive (data not
shown). Understanding this interesting case, where at comparable truncations
$w_s$ and $\kappa_1$ seem to grow differently, remains an open problem. We
mention that some scenarios leading to a non-exponential decay have been
recently discussed in Refs.~\cite{masud,sarang25} though for apparently
different reasons than those at play here.

\section{Generic quantum circuits}
\label{app:circs}

\begin{figure*}
    \centerline{\includegraphics[width=\textwidth]{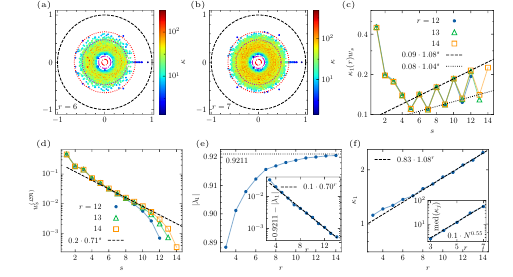}}
    \caption{\textbf{RP resonances and the fractal cascade in generic brickwall
    circuits}. The circuit used is homogeneous in space and time with the gate chosen randomly according to the Haar distribution \difa{($\vv \approx 1.1$)}. (a, b) Spectra of $\Mr$ for two values of $r$. Dotted
    lines show the single ring radii~\eqref{eq:sr_r}. The red radii (smaller
    inner circle, larger outer circle) are obtained from singular values of
    $\Mr$ for even $r$ and magenta radii (larger inner circle, smaller outer
    circle) from singular values of $\Mr$ for odd $r$. (c) Diverging partial norms~\eqref{eq:ws} of the
    normalized leading right eigenvector $\ket{\R_1}$ of $\Mr$. (d)
    Partial binorms~\eqref{eq:bin} of the leading left and right eigenvectors of
    $\Mr$ decaying exponentially. Left and right eigenvectors are binormalized,
    $\braket{\L_1}{\R_1} = 1$. (e) Convergence of the leading RP resonance
    $\lambda_1$. Convergence is exponential with a rate similar to the rate of decay
    of partial binorms. (f) The leading condition number $\kappa_1$ diverging
    with the same rate as the fastest divergence of partial norms. The $y$ axis
    is on a logarithmic scale.}
    \label{fig:generic}
\end{figure*}

In this appendix, we present numerical data for the quantities studied in the
main text for generic quantum circuits that are not dual-unitary. The circuits we are considering act on qubits
with the brickwall propagator defined in Eq.~\eqref{eq:circ_prop}, where the
gate $V$ is the same everywhere (i.e., the same $V_{i, j}$ for every $i, j$). To
obtain a generic representative, $V$ is chosen randomly according to the Haar
measure on the group of unitary $16 \times 16$ matrices. As we show in
Fig.~\ref{fig:generic}, the results follow all the predictions made in the main
text, with an additional staggering in $w_s$ on odd/even support. In the following paragraphs,
we briefly comment on the result.

The leading eigenvector $\ket{\R_1}$ of $\Mr$ exhibits the operator cascade
described in Sec.~\ref{sec:div_vec}, but with odd and even partial norms $w_s$
diverging with different rates, as depicted in Fig.~\ref{fig:generic}(c). We
find that generically, the odd-$s$ partial norms diverge more slowly than the
even-$s$. The staggering is not unexpected because the unit cell for brickwall circuits consists of two sites. The leading condition number $\kappa_1$
diverges with the larger of the two rates, as seen in Fig.~\ref{fig:generic}(f).
This guarantees that overlaps between the leading eigenvector and local
observables remain finite in the $r \to \infty$ limit, as discussed in
Sec.~\ref{sec:div_vec}.

\dif{The butterfly velocity of the considered circuit is approximately $\vv
\approx 1.1$ (determined from $w_s$ of a random linear combination of support-1 operators
evolved up to time $t = 4$ -- the largest possible before hitting a boundary of a finite system
on $L = 14$ sites), being close to the random-circuit result of $\vv = 6/5$~\cite{curt18,adam18}. The asymptotic equality \eqref{eq:ineq} holds well for the
divergence rate of the condition number $\mu \approx 1.08$ (the larger of the two
rates of divergence of partial norms); we obtain $\abs{\lambda_1} \mu^{1.1} \approx 1.004$.}

Furthermore, similarly as in the kicked Ising model (see Sec.~\ref{sec:conv}),
partial binorms of the leading left and right eigenvectors decay exponentially, as
shown in Fig.~\ref{fig:generic}(d). The leading RP resonance also converges
exponentially with a similar rate, as depicted in Fig.~\ref{fig:generic}(e). The
staggering is much less pronounced here, although some is visible,
particularly in the convergence of the resonance.

The spectrum of $\Mr$ exhibits a separation between the bulk and the RP
resonances similar to that in the kicked Ising model (see Sec.~\ref{sec:kappa}). We
now have different singular values of $\Mr$ on even and odd $r$, as opposed to
the kicked Ising model, where the singular values are the same for all $r$. We
conjecture that this is what one would generically see in unitary brickwall
circuits that are homogeneous in space and time. Different singular values now
result in two pairs of single ring radii~\eqref{eq:sr_r}; however, the bulk of
the eigenvalues seems to change relatively smoothly and does not stagger (i.e.,
conform to the prediction of the ``odd annulus'' for odd $r$ and ``even
annulus'' for even $r$). As seen in Fig.~\ref{fig:generic}(a, b), the bulk of
the eigenvalues is roughly localized in the vicinity of the two annuli for odd
and even $r$. Numerical data, however, is much less clean than in the kicked
Ising model. It is not clear whether one or the other annulus describes the bulk
better -- empirical results suggest it varies with the choice of gate $V$.

Finally, the spectrum also exhibits a separation between the bulk and the RP
resonances in the condition number $\kappa$~\eqref{eq:kappaj}, as observed in
Fig.~\ref{fig:generic}(a, b). Median $\kappa$ in the bulk diverges as $\sim
N^{0.55}$, where $N$ is the dimension of the Hilbert space, as seen in the inset
of Fig.~\ref{fig:generic}(f). The rate of divergence is similar to the $\sim \sqrt{N}$ found in the
kicked Ising model (see Sec.~\ref{sec:kappa}).

While we show results only for one gate realization, we tested the predictions on
many different circuits. The main difference between them is the value
of the leading RP resonance $\lambda_1$, the magnitude of which seems to
typically range from $0.4$ to $0.95$ (for additional examples see
Ref.~\cite{urban}).

\dif{

  \section{Conserved energy}
  \label{app:pre}

In the main text, we have analyzed systems with a $U(1)$ symmetry where the total
magnetization is conserved. One could argue that conservation of a quantity that
has its density supported on a single site, like magnetization, is qualitatively
different from the conservation of operators with density supported on more sites,
an example being energy. We numerically demonstrate here that conclusions
presented for $U(1)$ symmetry remain valid also for the conservation of energy.

A simple way to introduce conservation of energy in Floquet systems, i.e.,
quantum circuits, is to simply take the Trotter limit of an infinitesimally small
gate duration $\tau$. Taking our 1D kicked Ising model as an example, in the
limit $\tau \to 0$ the model will go towards the conservative Ising model in a
tilted field with Hamiltonian $H=\sum_j \sz_{j}\sz_{j+1}+\hz \sz_{j}+\hx
\sx_{j}$, which is generically chaotic. In order to be able to use our standard
truncated propagator formalism written for gates (Sec.~\ref{sec:Trunc}), one can
take small but finite $\tau$. For such small $\tau$ the energy is conserved only
approximately, i.e., up to some very long time, which, though, is not a problem
provided one is interested in physics at shorter times. On a numerical side, in
such a small-$\tau$ approach one must be aware that the gap will be small, and
therefore numerically finding $\lambda_1$ may be harder, and that the growth of
$\kappa_1$ and $w_s$ will also be correspondingly small ($\mu$ is only slightly
larger than $1$; beware though that this small growth is per one step of length
$\tau$, so the growth per unit of time will be $1/\tau$ times larger). Such an
approach to Hamiltonian systems has already been successfully used
before~\cite{prx}. In fact, because for Floquet systems one will typically
encounter prethermalization for small $\tau$~\cite{prethermalization}, we do not
even have to take a very small $\tau$ in order to (almost) conserve the energy.
Such prethermalization approach to Hamiltonian evolution has been also used in a
recent experiment on a quantum computer~\cite{feig25}. Choosing the kicked Ising
model for the same $\hx=0.9$ and $\hz=0.8$ as in the main text, but with
$\tau=0.45$, one already has a very small gap $1-\lambda_1 \approx 1.36 \cdot
10^{-5}$ (Fig.~\ref{fig:prethermalization}). Therefore, up to times of the order
of $t \sim 10^5$, there is a conserved operator that is equal to the tilted field
Ising Hamiltonian plus higher-order corrections (in $\tau$), see
Refs.~\cite{RP24,prx} for details.
\begin{figure}[t!]
  \centerline{\includegraphics[width=2.9in]{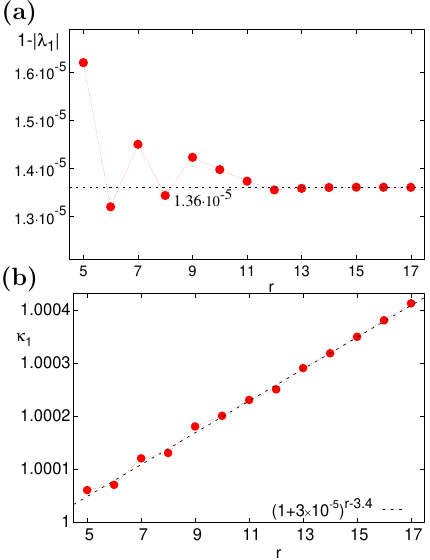}}
  \caption{\dif{{\bf Conserved energy.} 1D kicked Ising model in a prethermalization regime ($\tau=0.45$) where an almost conserved energy operator results in a small gap in the sector with $k=0$ (top), which in turn causes small growth of $\kappa_1$.}}
  \label{fig:prethermalization}
\end{figure}
While this operator will eventually decay, for our purpose of demonstrating
the diverging nature of eigenvectors at the studied truncations $r$ this long timescale
($\approx 10^5$) is irrelevant. Also, because the growth is very small due
to the small gap, it is better to look at the divergence of the condition number
$\kappa_1$ rather than at partial norms. We can see in
Fig.~\ref{fig:prethermalization} that the small growth rate $3 \cdot 10^{-5}$ is
compatible with an equally small RP gap (for a very small gap $\Delta=1-|\lambda_1|$
one expects $\mu\approx 1+\Delta/\vv$, and at $\tau=0.45$ one has $\vv \approx
0.5$). In contrast to the main text, where we focused on the nonzero
quasimomentum $k$, here the data is in the $k=0$ sector. Behavior in $k$ sectors
with nonzero small $k$ would be similar to the one for $U(1)$ conservation
presented in the main text, and would, in fact, be numerically easier to see due
to a larger gap.

\section{Scaling for the 2D kicked Ising model}
\label{app:2D}

In 1D systems, the leading RP resonance condition number diverged with the
truncation size $r$ as $\kappa_1 \sim \mu^r= (1/|\lambda_1|^2)^{\frac{r}{2\vv}}$, Eq.~\eqref{eq:kappalam}. That is,
divergence was exponential in $r$.

In Sec.~\ref{sec:2D}, we argued that in $d$-dimensional systems, the divergence should
be exponential in $\ell = r^{1/d}$, that is, in the linear size $\ell$ of the
truncated support and not in the total number of sites $r$ kept in truncation.
For a 2D system one should therefore have $\kappa_1 \sim (1/|\lambda_1|)^{a
\sqrt{r}}$. We have already shown in Fig.~\ref{fig:2Dkappa1} that the data for
the 2D kicked Ising model agrees with that expression well. However, because we
have data only for a limited range of $r$, perhaps some other dependence, for
instance $\kappa_1 \sim f^{ar^\alpha}$ with $\alpha \neq 1/2$, could fit equally
well. To that end, we plot in Fig.~\ref{fig:2Dscaling} the dependence of
$\ln\ln{\kappa_1}$ on $r$, from which one can infer the power
$\alpha$ from the slope.
\begin{figure}[t!]
  \centerline{\includegraphics[width=2.9in]{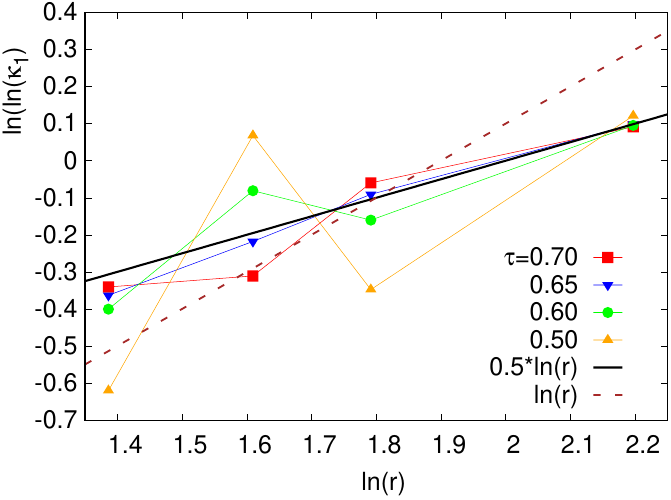}}
  \caption{\dif{{\bf Scaling exponent in 2D.} For the 2D kicked Ising model (\ref{eq:2D}) we plot the same data points as in Fig.~\ref{fig:2Dkappa1}. The black line indicating exponent $\alpha=\frac{1}{2}$ fits better than the dashed brown line with $\alpha=1$, see text for details.}}
  \label{fig:2Dscaling}
\end{figure}
We can see that the exponent $\alpha=\frac{1}{2}$ agrees with the data better than
$\alpha=1$, particularly if we exclude data for $\tau=0.50$ where
``fluctuations'' between different truncations are larger. This can be
understood as a consequence of $\tau=0.50$ already being close to
the prethermalization regime where longer timescales require larger truncation
sizes in order to get converged results (at $\tau=0.50$ the leading RP eigenvalue is $\lambda_1
\approx 0.78$), see Ref.~\cite{prx} where those
prethermalization effects have been analyzed in more detail.
}

\section{Matrix structure of the truncated propagator}
\label{app:structure}

The (truncated) propagator of the considered models is not a dense matrix due to
the sharp light cone. If one arranges the basis by increasing support and groups
the basis elements of each support in one block, it is a block-banded matrix,
where blocks below the diagonal describe the spreading of operators, and blocks
above the diagonal describe the shrinking of operators. In general, the number
of nonzero block bands is $4v + 1$. For simplicity, we additionally group $2v$
consecutive supports together, which allows us to represent $\M$ as a
block tridiagonal matrix
\begin{align}
    \M = \begin{bmatrix}
    \mathcal A_0 & \mathcal C_1 \\
    \mathcal B_1 & \mathcal A_1 & \mathcal C_2 \\
    & \mathcal B_2 & \mathcal A_2 & \ddots \\
    && \mathcal B_3 & \ddots \\
    &&&\ddots
    \end{bmatrix}. \label{eq:tridiag_u}
\end{align}
Matrix $\mathcal A_i$ is a square matrix acting in the space of operators with
support from $r = 2v i + 1$ to $r = 2v (i + 1)$, $\mathcal B_i$ is a rectangular
matrix mapping from the space of $\mathcal A_{i - 1}$ to the space of $\mathcal
A_i$, and $\mathcal C_i$ is a rectangular matrix mapping from the space of
$\mathcal A_i$ to the space of $\mathcal A_{i - 1}$. The size of blocks scales
exponentially as $\sim 4^{r}$.

Matrices $\mathcal A, \mathcal B, \mathcal C$ cannot be arbitrary, since
unitarity in the TDL $\mathcal U_{r \to \infty} \mathcal U_{r \to
\infty}^\dagger = \mathcal U_{r \to \infty}^\dagger \mathcal U_{r \to \infty} = \1$
enforces relations between them. In particular, using these relations,
one can see that for finite $r$, $\Mr \Mr^\dagger$ and $\Mr^\dagger \Mr$ deviate
from $\1$ only in the last block. Note that the deviation is still mathematically
large, since the size of the blocks scales exponentially, though physically
small as it changes the values of RP resonances exponentially little
(exponential convergence of $\lambda_j$ and binorms).

\subsection{Perturbation theory in increasing support}
\label{app:pert}

In Sec.~\ref{sec:conv}, we heuristically argued for the exponential convergence
of RP resonances. In this appendix, we work towards making the argument more
precise by employing perturbation theory. Our approximation, however, is
mathematically not well-controlled because, as mentioned, the perturbation is
small only on an appropriate physical space.

Let us consider increasing support from $r$ to $r + 2v$. This involves solving
the eigenproblem for the following matrix
\begin{align}
    \mathcal{U}_{r + 2v} = \left[
        \begin{array}{c | c} \Mr & \begin{matrix} 0 \\ \vdots \\ 0 \\
        \mathcal{C}_{i+1} \end{matrix} \\ \hline \begin{matrix} 0 & \dots & 0 &
        \mathcal{B}_{i+1} \end{matrix} & \mathcal{A}_{i+1} \end{array} \right],
\end{align}
given we already know the solution to the eigenproblem of $\Mr$, i.e., its
leading eigenvalue $\lambda_1^{(r)}$ and corresponding left and right
eigenvectors $\bra{\mathfrak L^{(r)}}$ and $\ket{\mathfrak R^{(r)}}$.
We denoted $i = r/(2v)$ and assumed that $r$ is a multiple of $2v$
for simplicity. First, we relabel the whole top right and bottom left block and
introduce a constant $\varepsilon$ that will help in keeping track of orders of
corrections
\begin{align}
	\mathcal{U}_{r + 2v} = \begin{bmatrix}
		\Mr & \varepsilon \mathcal C \\
		\varepsilon \mathcal B & \varepsilon \mathcal A
	\end{bmatrix}.
\end{align}
Here $\mathcal B = \mathcal B_{i + 1} \mathcal P_{i}$ and $\mathcal C =
\mathcal P_{i} \mathcal C_{i + 1}$, where $\mathcal P_{i}$ is the projector
to the basis of the block $i$ in Eq.~\eqref{eq:tridiag_u} (should not be
confused with the eigenprojector used in the main text).

We now decompose the vector into the two spaces defined by the block structure of
$\mathcal{U}_{r + 2v}$. Writing
$\ket{v} = \begin{bmatrix}
	\ket{v_1} \\
	\ket{v_2}
\end{bmatrix}$,
the eigenvalue equation is
\begin{align}
    \begin{bmatrix}
		\Mr & \varepsilon \mathcal C \\
		\varepsilon \mathcal B & \varepsilon \mathcal A
	\end{bmatrix} \begin{bmatrix}
		\ket{\mathfrak R^{(r + 2v)}_1} \\
		\ket{\mathfrak R^{(r + 2v)}_2}
\end{bmatrix} = \lambda^{(r + 2v)} \begin{bmatrix}
		\ket{\mathfrak R^{(r + 2v)}_1} \\
		\ket{\mathfrak R^{(r + 2v)}_2}
\end{bmatrix}. \label{eq:matrix_evq}
\end{align}
Now we express $\ket{\mathfrak R^{(r + 2v)}_2}$ from the second row of Eq.~\eqref{eq:matrix_evq}
\begin{align}
	\ket{\mathfrak R^{(r + 2v)}_2} = (\lambda^{(r + 2v)} - \varepsilon \mathcal A)^{-1} \varepsilon \mathcal B \ket{\mathfrak R^{(r + 2v)}_1} \label{eq:ev_ex}
\end{align}
and substitute it into the first row, obtaining
\begin{align}
    (\Mr + \varepsilon^2 \mathcal C (\lambda^{(r + 2v)} - \varepsilon \mathcal A)^{-1} \mathcal B) \ket{\mathfrak R^{(r+2v)}_1} = \lambda^{(r + 2v)} \ket{\mathfrak R^{(r+2v)}_1}. \label{eq:prs}
\end{align}

Eq.~\eqref{eq:prs} has the form of the standard Rayleigh-Schr\" odinger
perturbation theory, with the perturbation being equal to $\varepsilon^2
\mathcal C (\lambda^{(r + 2v)} - \varepsilon \mathcal A)^{-1} \mathcal B$. We
assume that the leading eigenvalue is non-degenerate and approximate
$\lambda^{(r + 2v)} \approx \lambda^{(r)}$ on the LHS. The standard first-order
perturbation theory now gives
\begin{align}
    \Delta \lambda(r) &= \lambda^{(r + 2v)} - \lambda^{(r)} \nonumber \\
                    &= \varepsilon^2 \bracket{\mathfrak L^{(r)}}{\mathcal C (\lambda^{(r)} - \varepsilon \mathcal A)^{-1} \mathcal B}{\mathfrak R^{(r)}} \nonumber \\
                   &= \frac{\varepsilon^2}{\lambda^{(r)}} \bracket{\mathfrak L^{(r)}}{\mathcal C \mathcal B}{\mathfrak R^{(r)}} + \mathcal{O}(\varepsilon^3) \label{eq:corr_vecs}\\
                   &= \frac{\varepsilon^2}{\lambda^{(r)}}\bracket{\mathfrak L^{(r)}}{\mathcal P_{i} \mathcal C_{i + 1} \mathcal B_{i + 1} \mathcal P_{i}}{\mathfrak R^{(r)}} + \mathcal{O}(\varepsilon^3), \label{eq:pert_res}
\end{align}
where we kept only the first non-trivial order in $\varepsilon$. The $\mathcal
O(\varepsilon^3)$ order can also be obtained by keeping the next order in
the Taylor expansion of $(\lambda^{(r)} - \varepsilon \mathcal A)^{-1}$. Further
higher orders, however, would also contain contributions from all the other
eigenvectors of $\Mr$.

Similarly, one can obtain the correction to the right and left eigenvectors
from Eq.~\eqref{eq:ev_ex}
\begin{align}
    \ket{\mathfrak R^{(r + 2v)}_2} &\approx \frac{\varepsilon}{\lambda^{(r)}} \mathcal B \ket{\mathfrak R^{(r)}}, \\
    \bra{\mathfrak L^{(r + 2v)}_2} &\approx \frac{\varepsilon}{\lambda^{(r)}} \bra{\mathfrak L^{(r)}} \mathcal C.
\end{align}
One can now use this in Eq.~\eqref{eq:corr_vecs} to connect the correction of
the eigenvalue to the correction of partial binorms
\begin{align}
	\Delta \lambda(r) &\approx \lambda^{(r)} \braket{\mathfrak L^{(r + 2v)}_2}{\mathfrak R^{(r + 2v)}_2} = \lambda^{(r)} \sum_{s = r + 1}^{r + 2v} w_s^{(\L \R)}.
\end{align}
The obtained result agrees with the prediction we argued for at the end of
Sec.~\ref{sec:conv}. An important caveat is that the perturbation theory we
employed is not well controlled. There is no small parameter $\varepsilon$, i.e., one
can numerically check that the operator norms of $\mathcal C \mathcal B$ are
of order $1$ in the considered systems. This means that we have no good argument
why higher-order terms contribute less than the term derived in this Appendix.
However, we can numerically check that the predicted corrections are of the
correct order of magnitude.


\begin{thebibliography}{99}
\bibitem{foot1} A stable (unstable) manifold \dif{of a fixed point} is a set of points that \dif{in $t \to \infty$ ($t \to -\infty$) converges to the fixed point}.
\bibitem{foot_unit} There is, of course, in principle no
contradiction between the unitarity of $\M$ and exponential relaxation, keeping in
mind (i) that one will have relaxation only for appropriate $A$, e.g., local
observables, and (ii) that one has to carefully take the correct TDL of first sending
$L \to \infty$ and only then $t \to \infty$, and in this case exponential decay
will indeed emerge out of exponentially many oscillating terms $\lambda_j^t$
on the unit circle, $|\lambda_j|=1$.
\bibitem{footP} \dif{It is not an orthogonal projector because $\P_j \neq \P_j^\dagger$. That is, denoting by $Y:={\cal R}(\P_j)$ the range of $\P_j$, one cannot split the Hilbert space into ${\cal H}=Y \oplus Y^\perp$, such that $\P_j(Y^\perp)=0$ because ${\rm ker}(\P_j)\neq Y^\perp$.}
\bibitem{foots} For instance, an operator with a density $a_1^{\ba}=\sx_1\sy_2$ has $s=2$, while $\sz_1\1_2\sx_3$ has $s=3$.
\bibitem{footws} \dif{Note that $w_s$ is the total weight of all operators with support $s$. Because their number grows as $\sim 4^s$, the weight of an individual component can decay even if $w_s$ grows (this happens if $\mu<4$).}
\bibitem{foot2} The inequality holds for diagonalizable matrices and eigenvalues with an algebraic multiplicity $1$. One can also define a condition number of a diagonalizable matrix $M=W D W^{-1}$ (with diagonal $D$) by $\kappa(M):= \nor{W}_p\cdot \nor{W^{-1}}_p$, and the Bauer-Fike theorem puts an upper bound on the spectral variation for $V$ of any size, i.e., difference in any two eigenvalues of $M$ and $M+V$, $|\lambda_i(M)-\lambda_j(M+V)| \le \kappa(M) \nor{V}_p$.
\bibitem{foot_du_align} The truncated propagator $\Mr$ in $k = 0$ for DU
circuits is actually equal to the non-translationally invariant propagator
aligning left-moving operators to the left of the causal cone and right-moving
operators to the right of the causal cone. On the level of the graph, the choice
of the direction of the shift $\mathcal T$ does not matter, since we are only interested
in the parity of the first site on which operators act nontrivially.
\bibitem{foot_dugr} The graph in Fig.~\ref{fig:du_graph} is valid for $r > 4$, but
the structure of fixed points is the same also for smaller $r$. All arguments are valid in the
general case -- large $r$ is considered merely for simplicity. For the complete
graph see Ref.~\cite{holden_dye}.
\bibitem{foot_dutm} The structure at $r_0$ must be calculated numerically, either
by diagonalizing $\Mr$ or the transfer matrices defined in Ref.~\cite{bruno}.
\bibitem{foot_duqc} At this point we assume that $\abs{\lambda_1} < 1$. For
$\abs{\lambda_1} = 1$, $\ket{\R_1}$ describes a conserved quantity localized in
$r_0$. Trivially, there is no cascade in this case. For a more general
discussion of the structure of conserved quantities see Ref.~\cite{holden_dye}.
\bibitem{footpsU} For normal operators, like a unitary or a Hermitian operator, the pseudospectrum and the spectrum coincide~\cite{trefethen}. The pseudospectrum of ${\cal U}$ is therefore again on a unit circle, i.e., is very different than the spectrum of $\Mr$.
\bibitem{footB} \dif{We note that under certain conditions one can prove~\cite{dolgopyat} that an exponential decay of correlations implies the Bernoulli shift property. The opposite, however, is not necessarily true. An example is the Bunimovich stadium~\cite{bunimovich}, which is chaotic but can have a power-law decay of correlations due to bouncing-ball orbits.}
\end{thebibliography}
\end{document}